\documentclass[conference]{IEEEtran-NDSS}

\usepackage{cite}

\ifCLASSINFOpdf
  \usepackage[pdftex]{graphicx}
\else
\fi

\usepackage{tikz}
\usepackage{amsmath,amssymb}
\usepackage{algorithm}
\usepackage{algpseudocode}
\usepackage{array}
\usepackage{fixltx2e}
\usepackage{stfloats}
\usepackage{url}
\usepackage{xcolor}
\usepackage{threeparttable}
\usepackage{makecell}
\setcellgapes{1.5pt}

\usepackage{amsfonts}
\usepackage{hyperref}

\usepackage[nameinlink]{cleveref}
\crefformat{section}{\S#2#1#3}
\Crefformat{figure}{#2Fig.~#1#3}
\crefname{figure}{Fig.}{Figs.}

\usepackage[shortcuts]{extdash}

\usepackage[caption=false,font=small]{subfig}
\usepackage[tablename=Table]{caption}

\usepackage{booktabs}
\usepackage{multirow}

\usepackage{microtype}

\usepackage{marvosym}

\ifdefined\mintedfinalize
\usepackage[finalizecache=true,cachedir=minted-cache]{minted}
\else
\usepackage[frozencache=true,cachedir=minted-cache]{minted}
\fi
\usepackage{inconsolata}
\usepackage{soul}
\usepackage{csvsimple}
\usepackage{ifthen}
\newcommand{\err}[1]{{\fontsize{6pt}{7pt}\selectfont$\pm$\,#1}}
\usepackage[normalem]{ulem}

\newcommand{\coolname}{SparSEEty}

\newcommand{\relullama}[1]{\textsf{ReluLLaMA#1}}
\newcommand{\llama}[1]{\textsf{Llama#1}}
\newcommand{\opt}[1]{\textsf{OPT#1}}
\newcommand{\nemotron}[1]{\textsf{Nemotron#1}}
\newcommand{\gemma}[1]{\textsf{Gemma#1}}

\newcommand{\smallrelullama}[1]{{\sf\small ReluLLaMA#1}}
\newcommand{\smallllama}[1]{{\sf\small Llama#1}}
\newcommand{\smallopt}[1]{{\sf\small OPT#1}}
\newcommand{\smallnemotron}[1]{{\sf\small Nemotron#1}}
\newcommand{\smallgemma}[1]{{\sf\small Gemma#1}}

\definecolor{sparseetyred}{HTML}{D62728}
\newcommand*\circledr[1]{\tikz[baseline=(char.base)]{
    \node[shape=circle,inner sep=.7pt,fill=sparseetyred] (char) {\textcolor{white}{\scriptsize\sffamily #1}};}}
\newcommand*\circledw[1]{\tikz[baseline=(char.base)]{
    \node[shape=circle,draw,inner sep=.75pt,fill=white, line width=.7pt] (char) {\scriptsize\sffamily #1};}}

\begin{document}
\title{\coolname{}:\\Extracting Tokens from Sparsity-Exploiting LLM Serving Systems via Deterministic Side Channels}

\author{%
\IEEEauthorblockN{Yongwan Jo\IEEEauthorrefmark{1}, Jinyoung Park\IEEEauthorrefmark{1}, Euihyun Lee, Dokyung Song\IEEEauthorrefmark{2}}
\IEEEauthorblockA{\it Department of Computer Science\\Yonsei University}\\

}

\IEEEoverridecommandlockouts

\maketitle
\renewcommand{\thefootnote}{\fnsymbol{footnote}}
\footnotetext[1]{Equal contribution.}
\footnotetext[2]{Corresponding author.}
\renewcommand{\thefootnote}{\arabic{footnote}}

\begin{abstract}
\noindent
Modern large language models (LLMs) exhibit \emph{activation sparsity}, wherein only a subset of their neurons is activated for given input tokens.
Researchers have leveraged this property to optimize LLM serving systems by omitting weight accesses and computations pertaining to inactive neurons.
Unfortunately, however, such optimizations create input-dependent weight accesses, which can be leaked over side channels.

We present \coolname{}, a new token extraction attack that exploits input-dependent neuron weight accesses introduced by sparsity-exploiting LLM serving systems.
\coolname{} first constructs a neuron-activation oracle using neuron weight access side channels during LLM inference, and then inverts the activation traces to reconstruct the input tokens, forming an end-to-end token extraction attack.
We instantiate \coolname{} against an LLM serving system protected inside an Intel~TDX confidential virtual machine (CVM), addressing three key challenges: (i)~constructing a neuron-activation oracle using a combination of side channels exposed by CVMs, (ii)~reducing inference-time overheads of neuron activation monitoring for covertness,
and (iii)~accurately inverting partial binary activation traces back to tokens.
Our evaluation shows that \coolname{} can reconstruct both prompt and response tokens with consistently high BLEU scores ($>$0.95) across various models and datasets, while incurring monitoring overheads of 3.7\% to 7.2\%.

\end{abstract}

\IEEEpeerreviewmaketitle

\section{Introduction}

\begin{figure}[t]
\centering
\includegraphics[width=0.495\textwidth,trim=2.6em 1.9em 1.5em 1.3em,clip]{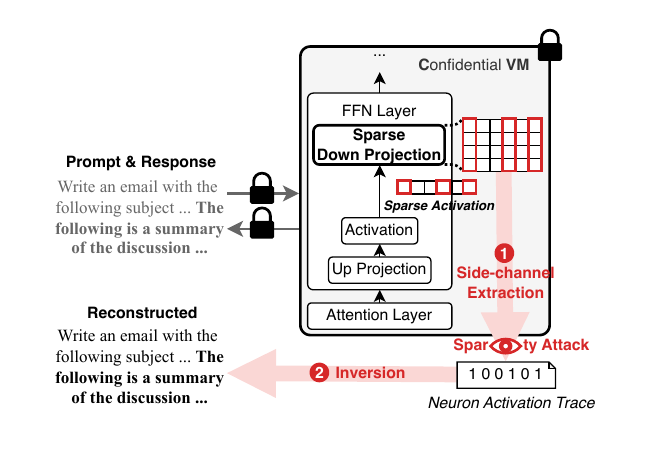}
\caption{High-level overview of the \coolname{} attack.}
\label{fig:idea}
\end{figure}

\noindent
Modern large language models (LLMs) exhibit \emph{activation sparsity}---a phenomenon in which only a subset of neurons is activated during inference depending on the input tokens.
This property offers both interpretability benefits, e.g., yielding more structured internal representations~\cite{elhage2022featuresuperposition} and improved model alignment~\cite{bayat2025sparseactivationsteering}, as well as significant efficiency gains:
Since only a small fraction of neurons in a Transformer~\cite{vaswani2017transformer} model's Feed-Forward Networks (FFNs) is activated in response to a given sequence of tokens, previous work has proposed to skip memory accesses and disk block accesses associated with inactive neurons~\cite{liu2023dejavu,zheng2023pit,alizadeh2024llminaflash}.
The resulting sparse computation significantly reduces both compute and memory overheads, enabling larger LLMs to be served with fewer resources~\cite{song2024powerinfer,xue2024powerinfer2,alizadeh2024llminaflash}.

However, this efficiency is gained at the cost of new privacy risks:
Namely, the memory access patterns resulting from sparse neuron computation become dependent on the input.
In many adversarial settings, these patterns can be leaked over side channels, e.g., via CPU cache monitoring~\cite{van2017withoutpagefaults}, page fault monitoring~\cite{xu2015controlledchannel}, or even physical address bus probing~\cite{zhuang2004hide,lee2020offchip,chuang2026teefail}.
Then, an attacker who observes these patterns could potentially correlate them with the input token sequence and infer sensitive information contained within it.

This paper concretely demonstrates the privacy risks of sparse neuron computation by presenting \coolname{}, an end-to-end token extraction attack against sparsity-exploiting LLM serving systems.
As illustrated in \Cref{fig:idea}, \coolname{} targets a Transformer-based LLM deployed inside a Confidential Virtual Machine (CVM), where users expect strong confidentiality guarantees for their prompts and generated responses.
In this setting, a user establishes a secure end-to-end communication channel with a CVM-hosted LLM service, and submits a prompt.
The prompt tokens, along with autoregressively generated tokens, pass through the Transformer layers, and each FFN performs sparse down projection based on sparse activation.
The generated tokens are returned back to the user as a response through the same secure channel.

The \coolname{} attack, operating under the standard CVM adversary model, proceeds in two phases.
First, during the LLM's inference on a victim's prompt, \coolname{} captures neuron activation patterns by monitoring selective accesses to down projection weights via a page fault side channel (also known as the controlled channel~\cite{xu2015controlledchannel}).
\coolname{} then inverts the captured neuron activation traces to reconstruct the tokens constituting the user's original prompt, as well as the LLM-generated response.

\coolname{} addresses three key challenges.
First, building a neuron-activation oracle from the memory-access oracle provided by page fault side channels requires identifying the guest physical addresses (GPAs) of each neuron's down projection weights, which reside in guest-private memory.
We achieve this by combining page fault side channels with two additional deterministic side channels: block I/O and page allocation side channels.
\coolname{} detects when the victim reads the target weights from its block device,
and then monitors subsequent accesses to recently allocated pages to efficiently establish the weight-to-GPA mappings.

Second, naive use of page fault side channels could incur costly CVM exits during victim inference, creating both performance overhead and detection risk.
We address this with two techniques: selective neuron monitoring and single-step-free chained page arming.
For selective monitoring, we apply an information-theoretic method~\cite{seo2024jointentropy} to identify a small subset of neurons that retain maximal discriminative power.
Our chained page arming technique further eliminates unnecessary exits caused by redundant page faults and single-stepping, while precisely capturing activation traces for all monitored neurons across all token positions.

Third, unlike prior inversion work~\cite{mahendran2015cnninversion,qu2025pia,dong2025depth} that assumes access to full-precision activation values, our oracle reveals only binary activations (i.e., activated or not), and only for a subset of neurons due to selective monitoring.
This creates ambiguity, as different tokens may produce identical binary patterns.
Moreover, a naive exhaustive search over the vocabulary at each token position is computationally expensive.
To address this, we guide the search using the probability distribution from a full forward pass of the LLM, conditioned on previously reconstructed tokens.
Candidate words are then tested in descending order of likelihood, and the first word whose binary activation pattern matches the observation is selected.
This approach significantly reduces the search overhead without compromising, and often even improving, reconstruction accuracy.

Our attack is related to (yet distinct from) prior work showing that intermediate activations can be used to reconstruct inputs.
Pioneered by Mahendran and Vedaldi~\cite{mahendran2015cnninversion}, who inverted intermediate representations of convolutional neural networks to reconstruct input images, recent studies have proposed inverting the activations of intermediate Transformer layers or even the final next token probabilities to reconstruct input tokens.
These methods, however, assume either access to full-precision activation values~\cite{qu2025pia,dong2025depth}---requiring the attacker to participate in inference or act as an auditor---or live query access to the victim LLM service for system prompt extraction~\cite{morris2024lminversion}.
In contrast, \coolname{} operates strictly under the standard CVM threat model: it extracts activation values at only binary precision through our neuron activation oracle, and requires no live query access to the target LLM service.

We implemented \coolname{} atop Linux KVM~\cite{kvm} and QEMU~\cite{bellard2005qemu}, targeting an LLM service running PowerInfer~\cite{song2024powerinfer} protected within an Intel~TDX CVM.
Our evaluation shows that \coolname{} can covertly extract neuron activation patterns with only 3.7\% to 7.2\% inference-time overhead across models, while accurately inverting the neuron activation traces.
Notably, monitoring only 100 first-layer FFN neurons, representing just 0.015\%--0.028\% of all FFN neurons across layers, suffices to accurately recover both prompt and response tokens (BLEU score $>$0.95) across all models and datasets.
\coolname{} remains effective even when a private LoRA~\cite{hu2021lora} adapter unknown to the adversary is applied, with a modest increase to 400 monitored neurons (still a small fraction of all FFN neurons).
More broadly, our findings expose a fundamental tension between efficiency and privacy in LLM serving:
as new resource-efficient inference optimizations emerge, their privacy risks demand rigorous investigation.

In summary, we contribute the following:

\begin{itemize}
\item We propose a new side-channel token extraction attack against CVM-protected LLM serving systems employing sparse computation for resource-efficient inference.
\item We propose a suite of techniques to (i)~extract neuron activation traces from a CVM using a combination of deterministic side channels, (ii)~reduce inference-time overhead caused by neuron activation monitoring for stealthy operation, and (iii)~invert partial, binary activation traces to accurately reconstruct both prompt and response tokens.
\item We implemented a complete prototype of our attack against an Intel~TDX-protected LLM serving system, and empirically demonstrate our attack can effectively recover the tokens while ensuring covert operation.
\end{itemize}

\section{Background}

\subsection{Using Activation Sparsity in LLM Serving}
\label{sec:background:llmserving}
\noindent
\textbf{Transformer-based LLMs} take as input a sequence of tokens (i.e., an input prompt) denoted by $(x_1, \dots, x_n)$, where each token $x_i$ is drawn from a fixed vocabulary $\mathcal{V}$,
and generate output tokens $(x_{n+1}, \dots, x_{n+T})$ during inference~\cite{vaswani2017transformer}.
These output tokens are generated autoregressively, one token at a time, until the model emits a special end-of-sequence (EOS) token (typically $x_{n+T}$).
LLM inference proceeds in two stages: prefill and decode.
In the prefill stage, the model processes the entire input prompt to compute a probability distribution $P(v | x_1, \dots, x_n) : v \in \mathcal{V}$, from which it samples the first new token $x_{n+1}$.
In the decode stage, the model computes the probability distribution for each subsequent token autoregressively, conditioning on the input prompt and all previously generated tokens: $P(v | x_1, \dots, x_n, \dots, x_{n+i-1}) : v \in \mathcal{V}$.
The model then samples $x_{n+i}$ from this distribution, and the process continues until an EOS token is generated.

These LLMs stack multiple Transformer layers, each transforming input representations to output representations of the same size, referred to as the hidden dimension $d_{\textnormal{model}}$.
Each Transformer layer includes a Feed-Forward Network (FFN) sublayer,
a two-layer multi-layer perceptron comprising two linear transformations with an activation function, e.g., Rectified Linear Unit (ReLU), applied between them.
The FFN first applies \emph{up projection}, a linear layer that maps the input to a higher-dimensional space (from $\mathbb{R}^{d_{\textnormal{model}}}$ into $\mathbb{R}^{d_{\textnormal{ff}}}$).
The result is passed through an activation function, and then projected back to $d_{\textnormal{model}}$ by another linear layer called \emph{down projection}.

\smallskip
\noindent\textbf{Dynamic Activation Sparsity}
refers to a phenomenon where neurons in the Transformer's FFN sublayers are selectively activated depending on the current input~\cite{li2023transformersparsity}.
The degree of sparsity varies across models, with each neuron activated for only 1.48\% to 25.65\% of tokens on average in ReLU and ReLU$^2$ models, according to our profiling (see \Cref{sec:evaluation}).

\begin{itemize}
\item
\noindent\textbf{Naturally-Occurring Activation Sparsity.}
For models using ReLU (or its variants such as ReLU$^2$~\cite{zhang2024relusquared}) as their activation functions, a large degree of activation sparsity naturally arises as any negative values are mapped to zero after applying the activation function.

\item
\noindent\textbf{Artificially-Induced Activation Sparsity.}
For models employing activation functions other than ReLU variants, the proportion of zero-valued activations is typically negligible.
Prior work has proposed to artificially induce dynamic activation sparsity, either by \emph{ReLU-fying} the model ahead of time (i.e., replacing the activation function with ReLU followed by fine-tuning)~\cite{mirzadeh2024relufication} or by applying run-time thresholding to suppress low-magnitude activations~\cite{federici2025dip}.
\end{itemize}

\smallskip
\noindent\textbf{Sparse Down Projection in FFN Sublayers.}
Prior work leveraged activation sparsity to perform the down projection computations in Transformer's FFN sublayers~\cite{liu2023dejavu,liu2025trainingfreesparsity} via sparse matrix-vector multiplication.
The core idea is that dynamic activation sparsity produces highly sparse activations within each FFN, and the columns of the down projection weight matrix corresponding to zero-valued activations (arising from ReLU-like functions and other activation functions followed by magnitude-based thresholding) can be skipped without affecting the output.
A major advantage of using sparse matrix-vector multiplication is the reduction in computational cost.
For instance, Mirzadeh et al. report a 32\% decrease in computation for \smallopt{-6.7b}~\cite{mirzadeh2024relufication}.
Researchers further proposed to predict which neurons will be activated before applying activation, enabling the sparse matrix-vector multiplications to be applied to the up projection as well~\cite{liu2023dejavu}.

\subsection{Deterministic Side Channels in CVMs}
\label{sec:background:sidechannel}
\noindent
We now describe deterministic side channels that we exploit in our token extraction attack.
We focus on the side channels that
arise in virtualization-based trusted execution environments (TEEs), also known as confidential virtual machines (CVMs).

\smallskip
\noindent\textbf{Page Access Side Channel.}
TEEs commonly delegate memory management to the untrusted host for reducing the complexity and size of the trusted computing base.
Exploiting this design choice, Xu et al.~\cite{xu2015controlledchannel} demonstrated page fault side channel attacks (also known as controlled-channel attacks) against Intel~SGX~\cite{intel-sgx},
in which the host OS removes virtual-to-physical page mappings to trigger enclave page faults.
Similar attacks have also been shown against CVMs~\cite{wilke2020sevurity,li2019exploitingamdsev,wilke2024tdxdown,hornetz2026tdxray}, including those built on AMD~SEV~\cite{amd-sev} and Intel~TDX~\cite{intel-tdx},
through manipulation of mappings from guest physical addresses (GPAs) to host physical addresses (HPAs).

In Intel~TDX, the trusted TDX module~\cite{intel-tdx-module-spec}, running in a newly added CPU mode called Secure Arbitration Mode (SEAM), manages Secure Extended Page Tables (SEPTs) that contain GPA-to-HPA mappings for CVM-private pages.
This module exposes a SEAMCALL interface to the untrusted host.
It includes the following functions that can be used by host-side adversaries to manipulate SEPT entries and trigger page faults in CVMs, thereby creating page fault side channels.

\begin{itemize}
\item \mintinline{bash}|TDH.MEM.RANGE.BLOCK| {blocks} guest access to specified GPA ranges, triggering page faults on subsequent accesses.

\item \mintinline{bash}|TDH.MEM.TRACK| increments the CVM's TLB tracking counter, which is used for TLB shootdown described below.

\item \mintinline{bash}|TDH.MEM.RANGE.UNBLOCK| {unblocks} specified GPA ranges previously blocked, restoring guest access to them.
\end{itemize}

When the host blocks guest pages, it must perform a TDX-specific TLB shootdown to invalidate stale GPA-to-HPA mappings cached on other cores executing the same CVM.
This is done by issuing inter-processor interrupts to force those cores to exit the CVM, and then invoking \mintinline{bash}|TDH.MEM.TRACK| to increment the global counter.
Each core maintains a local counter, and, upon detecting a mismatch with the global counter on its next entry into the CVM, flushes all CVM-associated TLB entries.
Once a GPA range is blocked, accesses trigger page faults without populating the TLB;
consequently, unblocking the range does not require a shootdown, and normal TLB caching for the unblocked range resumes.

\smallskip
\noindent\textbf{Page Allocation Side Channel.}
Intel~TDX supports demand paging for efficient memory management.
When a CVM accesses unallocated pages, it triggers a CVM exit with an SEPT fault.
The host handles this by invoking \mintinline{bash}|TDH.MEM.PAGE.AUG|, a SEAMCALL that requests additional private pages.
This creates a side channel: a malicious host can observe the timing and sequence of the CVM's page allocation.

\smallskip
\noindent\textbf{Block I/O Side Channel.}
CVMs rely on block devices provided by the untrusted host virtual machine monitor (VMM) such as QEMU~\cite{qemuuserguide}, for non-volatile storage that hosts their filesystems.
CVMs access these devices by requesting sectors to the VMM, specifying (i)~the sector offset and (ii)~the GPA for the destination.
This creates a block I/O side channel: an attacker on the host VMM side can observe (i)~which sectors in the block device are requested by the CVM, and (ii)~which GPAs receive them.
Notably, the GPA provided by the CVM points to a bounce buffer located in its non-private memory shared with the VMM, since the VMM is not permitted to access the CVM's private memory.
Once the VMM copies the requested sectors into the bounce buffer, the CVM copies (or ``bounces'') them back into its private memory when resumed, typically into their OS's page cache.

\section{Threat Model \& Assumptions}
\label{sec:threatmodel}
\noindent
We consider deploying an LLM service within a CVM, such as a Trust Domain in Intel~TDX, reflecting growing interest in deploying DNN serving within TEEs across both academia~\cite{mo2020darknetz,siby2024guarantee,moon2025asgard,tan2025pipellm,wang2026tzllm} and industry~\cite{google-private-ai,anthropic-confidential-inference,microsoft-confidential-ai}.
We focus on private, CPU-based inference scenarios where each user provisions a dedicated CVM hosting an LLM service to securely process sensitive prompts,
leveraging CPU TEEs as a secure and cost-effective alternative to GPU TEEs for moderate batch and input sizes~\cite{chrapek2025confidentialllminference}.

To improve computational efficiency, the victim LLM service is assumed to employ the following optimizations: (i)~model parallelism~\cite{shoeybi2019megatron}, where multiple CPUs are allocated to the CVM, and multiple threads are spawned within it to enable parallel execution of LLM inference,
(ii)~KV cache, which removes redundant computation caused by Transformer's attention mechanism during autoregressive token generation, by caching the key (K) and value (V) vectors of all Transformer layers for all prompt tokens and previously generated tokens,
and (iii)~sparsity-exploiting optimizations, specifically sparse matrix-vector multiplication in the FFN layers during the down projection step, leveraging dynamic activation sparsity (see \Cref{sec:background:llmserving}).
An optimized LLM serving framework called PowerInfer~\cite{song2024powerinfer}, for example, employs all of these optimizations.
In this setting, we assume a standard, host-side CVM attacker who aims to infer the tokens being processed by the victim LLM service, by exploiting side channels exposed by the CVM.

\smallskip
\noindent\textbf{Defensive Assumptions.}
We assume that the CVM exposes the LLM service to legitimate users only, via an authenticated, encrypted communication channel; in other words, the attacker has no query access to the victim LLM service.
This implies that prompt extraction attacks across co-tenants of the LLM service, such as PromptPeek~\cite{wu2025promptpeek}, are not applicable in our setting.
We also assume that known side-channel prompt extraction attacks, such as the one proposed by Gao et al.~\cite{gao2025iknowwhatyousaid}, which targets the victim LLM service's embedding table access, are mitigated.
This can be achieved by removing the embedding lookup accesses, e.g., through deep hash embedding techniques~\cite{kang2021deephashembedding,umar2025deephashembedding}.

We also assume a typical configuration of a CVM, which uses a standard set of \emph{virtio} devices~\cite{russel2008virtio,virtiov1.2}, including \emph{virtio-blk} and \emph{virtio-net} for providing block storage and network connectivity, respectively.
Because block devices are controlled by the untrusted host, we assume that the CVM employs a block device integrity protection mechanism, such as \emph{dm-verity}~\cite{google2024dmverity}, to prevent the host from tampering with the LLM weights stored on disk, which could otherwise be manipulated to facilitate prompt leakage.
We also assume that the LLM service and the kernel inside the CVM uses address space layout randomization, ASLR and KASLR, respectively.

\smallskip
\noindent\textbf{Adversarial Capabilities.}
We adhere to the standard, software attacker model for CVMs,
where the attacker controls the majority of privileged system software including the host VMM, but without physical access to hardware.
Specifically, we consider the attacker assumed by Intel~TDX~\cite{intel-tdx-google-review}, since our focus is on LLM services running inside an Intel~TDX CVM.
The attacker can observe sequences of memory pages and block sectors accessed by the CVMs, through TDX's page fault side channel and block I/O side channel (see \Cref{sec:background:sidechannel}), respectively.
In line with prior work~\cite{qu2025pia,luo2026kvinversion,gao2025iknowwhatyousaid}, we also assume that the attacker has full offline access to the LLM weights.
This captures common deployment scenarios for private LLM services, where the served model is either a publicly available base model or a variant adapted to private downstream tasks via parameter-efficient fine-tuning (PEFT), e.g., LoRA~\cite{hu2021lora}.
We do not, however, assume knowledge of the PEFT adapter weights, as these can be securely provisioned to the CVM.
When the base model is unknown, the attacker may first attempt model fingerprinting~\cite{pasquini2025llmmap} or stealing attacks~\cite{hua2018reversecnn,yan2020cachetelepathy,gao2024deeptheft,dong2025depth} via query APIs, side channels, or a combination of both;
we consider these directions orthogonal to our focus and leave them to future work (see \Cref{sec:discussion}).

\section{The \coolname{} Attack}
\label{sec:overview}
\noindent
We now present \coolname{}, a new token extraction attack against sparsity-exploiting LLM serving systems (see~\Cref{sec:background:llmserving}).
Our key insight is that these systems selectively access the down projection weights in an LLM's FFN layers depending on which of its neurons are activated, and that such access patterns can be leaked through side channels and then inverted to recover the tokens processed by the LLM.

\smallskip
\noindent\textbf{Notation \& Adversarial Goals.}
Let
$T^{(j)}(\cdot)$ denote the $j$-th Transformer layer.
We define $A^{(j)}(\cdot)$ as the portion of $T^{(j)}(\cdot)$ up to (and including) the activation function within it, whose output is fed into its down projection linear layer denoted by $D^{(j)}(\cdot)$.
Since page fault side channels expose activation traces in binary form, we let $B^{(j)}(\cdot)$ denote the binarized neuron activation trace of $A^{(j)}(\cdot)$.
We then define $\mathbf{B}( \cdot )$ as the concatenation of these binarized neuron activation traces across all Transformer layers.
For an input token $x_i$, omitting attention to previous tokens for simplicity, this is expressed as:
\begin{equation}
\label{eq:binarytrace}
\mathbf{B}(x_i) = B^{(1)}(x_i) \mathbin\Vert B^{(2)}(T^{(1)}(x_i)) \mathbin\Vert \dots
\end{equation}

Using this notation, the sequence of binary activation traces leaked by the victim LLM during inference is given as:
\begin{equation}
\label{eq:binarytracesequence}
\begin{split}
\Big\{\mathbf{B}(x_1), \dots, \mathbf{B}(x_{n}), \mathbf{B}(x_{n+1}), \dots, \mathbf{B}(x_{n+T-1})\Big\}
\end{split}
\end{equation}
The first $n$ activation traces are leaked during the prefill stage, as the LLM processes the input tokens $(x_1, \dots, x_n)$ to produce the first output token $x_{n+1}$.
The remaining $T-1$ activation traces are leaked during the decode stage, as the LLM generates subsequent output tokens, starting from $x_{n+2}$ to $x_{n+T}$.

\coolname{'s} goals can be formulated as follows: (i)~to extract the binary activation trace sequence given in \Cref{eq:binarytracesequence} covertly during the victim LLM's execution,
and (ii)~given these $n+T-1$ activation traces, to find a token sequence $(\hat{x}_1, \dots, \hat{x}_{n+T-1})$ that best matches the original sequence $({x}_1, \dots, {x}_{n+T-1})$.

\smallskip
\noindent\textbf{Attack Flow.}
\coolname{} operates in two phases: an online and offline phase.
The online phase (see \Cref{sec:tracing}) begins with the attacker identifying the GPAs to which the LLM's down projection weights are loaded (\Cref{sec:tracing:gpa}).
The attacker then arms those pages,
causing page faults to trigger as the victim LLM processes input prompts and generates output tokens (\Cref{sec:tracing:pagefault}).
The resulting faults are collected to construct a sequence of binary activation traces for each processed and generated token.
To reduce the number of CVM exits and page faults, \coolname{} also incorporates several optimizations (\Cref{sec:tracing:optimizations}).

In the offline phase (\Cref{sec:inversion}), the attacker inverts this sequence of neuron activation traces to reconstruct the original tokens.
Starting with the first neuron activation trace to reconstruct the first token, \coolname{} autoregressively inverts the neuron activation trace of each subsequent token (\Cref{sec:inversion:autoregressive}).
\coolname{} uses a search-based method to invert each activation trace (\Cref{sec:inversion:search}) to accurately and efficiently recover the original tokens.

\begin{table*}[t]
\caption{Comparison between token reconstruction attacks against LLM serving systems.}
\label{tab:inversion:comparison}
\centering
\footnotesize
\setlength\tabcolsep{0.14em}
\begin{threeparttable}
\begin{tabular}{|l|c|c|c|c|c|c|c|}
\hline
\makecell[cc]{Attack}
  & \makecell[cc]{Threat model}
  & \makecell[cc]{Leaked internal state}
  & \makecell[cc]{Observation channel}
  & \makecell[cc]{Query access}
  & \makecell[cc]{Weight access}
  & \makecell[cc]{Adversary}
  & \makecell[cc]{Reconstructed tokens}
  \\
\hline
\makecell[lc]{Language Model Inversion~\cite{morris2024lminversion}}
   & \makecell[cc]{Black-box API}
  & \makecell[cc]{Next-token probabilities}
  & \makecell[cc]{API response}
  & Required
  & None
  & \makecell[cc]{API client}
  & \makecell[cc]{System prompt}
  \\
\hline
\makecell[lc]{PIA~\cite{qu2025pia}}
  & \makecell[cc]{Collaborative\\inference}
  & \makecell[cc]{Intermediate activations\\(partial layers)}
  & \makecell[cc]{Direct tensor read}
  & None
  & Required
  & \makecell[cc]{Malicious\\participant}
  & \makecell[cc]{Prompt \& response}
  \\
\hline
\makecell[lc]{Dong et al.~\cite{dong2025depth}}
  & \makecell[cc]{Collaborative\\inference}
  & \makecell[cc]{Intermediate activations\\(partial layers)}
  & \makecell[cc]{Direct tensor read}
  & \makecell[cc]{Required}
  & \makecell[cc]{None}
  & \makecell[cc]{Malicious\\participant}
  & \makecell[cc]{Prompt \& response}
  \\
\hline
\makecell[lc]{I Know What You Said~\cite{gao2025iknowwhatyousaid}}
  & \makecell[cc]{Co-tenant on\\shared host}
  & \makecell[cc]{Embedding table\\lookups}
  & \makecell[cc]{Cache-line access\\pattern}
  & None
  & Required
  & \makecell[cc]{Co-located\\process}
  & \makecell[cc]{Prompt \& response}
  \\
\hline
\makecell[lc]{KV-cache Inversion~\cite{luo2026kvinversion}}
  & \makecell[cc]{Confidential\\LLM serving}
  & \makecell[cc]{Cached KV values}
  & \makecell[cc]{Direct tensor read}
  & None
  & Required
  & \makecell[cc]{Untrusted\\host}
  & \makecell[cc]{User prompt}
  \\
\hline
\makecell[lc]{TDXRay~\cite{hornetz2026tdxray}}
  & \makecell[cc]{Confidential\\LLM serving}
  & \makecell[cc]{Tokenization hash map\\lookups}
  & \makecell[cc]{Page/cache-line\\access pattern}
  & Required
  & \makecell{None}
  & \makecell[cc]{Untrusted\\host}
  & \makecell[cc]{Prompt \& response}
  \\
\hline
\makecell[lc]{\bf \coolname{} \bf (Ours)}
  & \makecell[cc]{Confidential\\LLM serving}
  & \makecell[cc]{Binary activations}
  & \makecell[cc]{Block/page\\access pattern}
  & None
  & Required
  & \makecell[cc]{Untrusted\\host}
  & \makecell[cc]{Prompt \& response}
  \\
\hline
\end{tabular}
\end{threeparttable}
\end{table*}

\smallskip
\noindent\textbf{Comparison with Prior Token Reconstruction Attacks.}
Prior work has investigated token reconstruction for LLMs across a range of settings, as summarized in \Cref{tab:inversion:comparison}.
These attacks differ along three key dimensions: their threat models, the information leaked and the channels through which it leaks, and the types of tokens targeted for reconstruction.

A line of work proposed to invert the final next-token probabilities of LLMs~\cite{morris2024lminversion}, intermediate activations~\cite{qu2025pia,dong2025depth}, and KV cache values~\cite{luo2026kvinversion} to reconstruct system or input prompts.
These attacks operate, respectively, in a black-box API setting (where the attacker directly observes API responses), a collaborative inference setting (where the attacker observes partial intermediate activations), and a confidential LLM serving setting (where the attacker observes KV cache offloaded to an untrusted host).
Another line of work exploits side channels to leak embedding table lookups~\cite{gao2025iknowwhatyousaid} or tokenization hash map lookups~\cite{hornetz2026tdxray} to reconstruct input and output tokens in a co-located process setting and a confidential LLM serving setting, respectively.
Our attack assumes a unique setting: the attacker is able to observe (a select subset of) intermediate neuron activations triggered by both prompt and response tokens through a combination of side channels, but, unlike prior work, only at binary resolution; that is, the attacker can determine whether each neuron was activated or not (a binary value), without access to the exact activation magnitude.

Existing attacks assume either query access or offline weight access, but never neither. When offline weight access is not assumed, query access is typically required to profile the victim model's behavior: Dong et al. use query access to train a surrogate model that mimics the victim LLM~\cite{dong2025depth}, and TDXRay uses it to profile the location of hash maps~\cite{hornetz2026tdxray}.
Our attack assumes offline weight access but not query (inference API) access.
We note, however, that, like prior work, query access could substitute for offline weight access by profiling the victim model's activation patterns over the attacker-chosen token sequences and then matching observed patterns against the profile to reconstruct the victim's tokens.

\section{Collecting Neuron Activation Traces}
\label{sec:tracing}

\noindent
The \coolname{} attack begins with an online phase, which takes place while the victim LLM serving system is processing tokens (hence the term ``online'').
During this phase, the attacker observes the victim system via a combination of block and page access oracles, and extracts a neuron activation trace for each token processed by the LLM.
\coolname{} extracts these traces by monitoring selective access patterns to the down projection weights of FFN neurons, arising from sparse activations followed by sparse down projection.

\begin{figure}[t]
\centering
\includegraphics[width=0.495\textwidth,trim=2.3em 1.1em 2.1em 0.9em,clip]{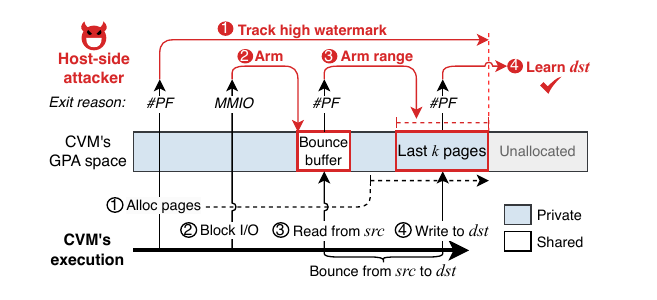}
\caption{Finding private GPAs of down projection weights.}
\label{fig:gpa}
\end{figure}

\subsection{Finding GPAs of Down Projection Weights}
\label{sec:tracing:gpa}
\noindent
The first step in collecting neuron activation traces during LLM inference is to identify the GPAs of down projection weights selectively accessed due to \emph{sparse} down projection.
To this end, we propose to combine block access oracle with page allocation and access oracles, as detailed below.

\Cref{fig:gpa} illustrates how we combine them to learn the GPAs of down projection weights.
During bootstrapping, the victim LLM system inside the CVM maps the model file into its address space, prompting the OS's block layer to allocate CVM-private pages to receive block data (see \circledw{1}).
When the CVM accesses previously unallocated pages, it triggers an SEPT page fault, which the attacker uses to track the high watermark of the CVM's GPA space (\circledr{1}).

Next, the CVM's {\it virtio-blk} driver issues block read requests (\circledw{2}).
\coolname{} monitors the requests that target the down projection weights, and learns
the GPA of the bounce buffer to which these weights are copied.
Selective weight accesses during LLM inference, however, do not occur on the bounce buffers; instead, they occur on the CVM-private pages within the CVM's OS page cache, where the weights are ultimately copied.
To this end, \coolname{} arms the bounce buffer (\circledr{2}) and resumes the CVM.

Once resumed and signaled of block I/O completion, the CVM reads the bounce buffer (denoted $src$) for bouncing (\circledw{3}), triggering an EPT fault.
Upon receiving this fault, \coolname{} disarms the bounce buffer, and arms the last $k$ pages below the high watermark (\circledr{3}), one of which is the final destination ($dst$) of the bounced data.
While \coolname{} could block all guest pages by setting $k$ to the total number of guest pages, doing so would incur excessive \mintinline{bash}{TDH.MEM.RANGE.BLOCK} SEAMCALLs.
For finding the GPAs of down-projection weights, we set $k = 16,000$ based on our empirical observation that the LLM's large size causes it to be loaded into pages near the high watermark.

\coolname{} then resumes the CVM, which immediately triggers another CVM exit with an SEPT fault, as the CVM attempts to write the bounced data to an OS page-cache page (\circledw{4}).
By examining the GPA of the faulting page, the attacker finally learns the GPAs where the down projection weights are ultimately stored in CVM-private memory (\circledr{4}).

\subsection{Extracting Activations via Page Faults}
\label{sec:tracing:pagefault}
\noindent
Having identified the GPAs of the CVM-private pages containing the down projection weights in the previous step,
we can now determine which neurons are activated in each FFN layer by inducing and monitoring page faults when the CVM reads these weights during sparse down projection.

\smallskip
\noindent\textbf{Observability of Per-Neuron Activation.}
An inherent limitation of controlled-channel attacks is that it reveals memory access patterns only at the granularity of pages.
In practice, however, this does not hinder \coolname{'s} ability to monitor neuron activations.
In modern LLMs, the hidden dimension size (i.e., $d_{\textnormal{model}}$), which determines the size of the down projection weights for a single neuron, typically reaches into the thousands, with each weight stored using 16-bit floating-point precision.
As a result, the down projection weights for a single neuron typically span an entire 4KiB page (or more) in memory.
For example, in \smallllama{-2-7b}~\cite{llama2}, the size of the down projection weights per neuron is 8KiB.
By monitoring either of the two 4KiB pages, \coolname{} can reliably determine whether each neuron was activated.

\smallskip
\noindent\textbf{Handling Model Parallelism.}
Modern LLM systems
employ model parallelism~\cite{shoeybi2019megatron},
where the down projection $D^{(j)}(\cdot)$ is split across multiple threads, which run across different virtual CPU (vCPU) cores in the case of CVMs.
Each thread simultaneously processes a subset of neurons during the down projection, and the partial results of projection are then reduced (i.e., summed) to produce the full matrix-vector multiplication result.
\coolname{} handles this model parallelism, by collecting concurrent page faults from different vCPUs and aggregating them for each token.

\smallskip
\noindent\textbf{Handling Batched Prefill Computation.}
During the prefill stage, LLM systems typically process the entire prompt $(x_1, \dots, x_n)$ as a batch, sending all tokens through each layer before moving on to the next.
For $D^{(j)}(\cdot)$, this means that the system computes $D^{(j)}(x_1)$, \dots, $D^{(j)}(x_{n})$ sequentially before $D^{(j+1)}(\cdot)$.
\coolname{} handles this batched computation by disentangling the per-token activation patterns within each layer, and then aggregating these layer-wise patterns across all Transformer layers to construct complete activation profiles for each position in the prompt token sequence.

\subsection{Optimizations for Enhancing Stealthiness}
\label{sec:tracing:optimizations}

\noindent
Using page faults for monitoring neuron activations could introduce significant overhead due to VM exits triggered by page faults.
This slows down the CVM's LLM inference, raising the likelihood of detection.
We therefore propose two optimizations to enhance \coolname{'s} stealthiness.

\smallskip
\noindent\textbf{Selective Neuron Monitoring.}
One could monitor every neuron in every FFN layer collecting the full binary activation trace $\mathbf{B}(x_i)$ for each token $x_i$, but this requires installing a prohibitively large number of page fault probes, resulting in a page fault for every activated neuron.
To mitigate this, we propose a selective neuron monitoring strategy.

\begin{itemize}
\item {\bf Layer Selection:} We first select activations of the lower layers of the victim LLM for monitoring, based on the finding that the activations of lower layers that are closer to the token embedding layer contain more direct information about the input~\cite{mahendran2015cnninversion}.
\coolname{} uses the activation trace of the first FFN layer for each token, i.e., $B^{(1)}(x_i)$ for $x_i$.
\item {\bf Neuron Selection:}
Within the first layer, we further select a subset of neurons whose combined activations maximize both information content and discriminative power~\cite{seo2024jointentropy}, as follows.
Using a public dataset $\mathcal{A}$, we generate binary activation traces $B^{(1)}(x)$ for all tokens $x \in \mathcal{A}$, forming the empirical distribution used to compute entropy.
We begin by selecting the neuron with the highest individual entropy, then iteratively expand the set via a greedy search, each time adding the neuron that yields the largest increase in joint entropy.
This yields a compact yet informative subset of neurons capable of distinguishing diverse activation patterns.
We denote the partial activation trace of this selected subset in the $k$th layer by $\tilde{B}^{(k)}( \cdot )$;
during inference, \coolname{} monitors $\tilde{B}^{(1)}(x_i)$ for each token $x_i$.
\end{itemize}

This two-stage neuron selection process of \coolname{} yields the following pruned sequence of activations when observing the token sequence $(x_1, ..., x_{n+T-1})$:
\begin{equation}
\label{eq:prunedbinarytrace}
\begin{split}
\Big\{\tilde{B}^{(1)}(x_1), \dots, \tilde{B}^{(1)}(x_{n}), \tilde{B}^{(1)}(x_{n+1}), \dots, \tilde{B}^{(1)}(x_{n+T-1})\Big\}
\end{split}
\end{equation}
We show later that monitoring this pruned set of neurons is sufficient for reconstructing the tokens with high accuracy.

\begin{figure}[t]
\centering
\includegraphics[width=0.495\textwidth,trim=2.2em 0.7em 1.55em 0.9em,clip]{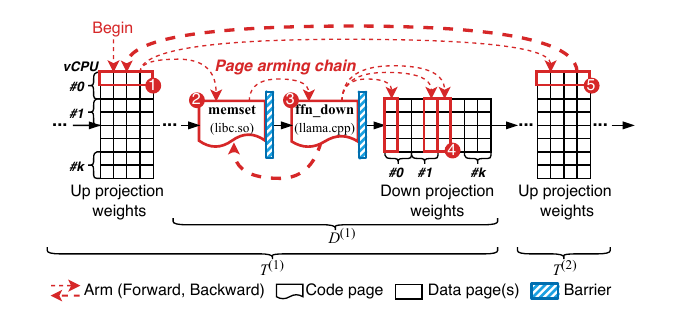}
\caption{Page-arming chain used to monitor the activations of the first FFN layer under multi-threaded, model-parallel LLM inference.}
\label{fig:zapping}
\end{figure}

\smallskip
\noindent\textbf{Single-Step-Free, Chained Page Arming.}
\noindent
A naive approach to monitoring memory accesses via page faults is to arm each monitored page, single-step the CVM on every VM exit caused by accessing an armed page, and re-arm the page upon another VM exit caused by single-stepping.
While using this approach ensures that every access to the down projection weights is monitored, it has two critical drawbacks: (i)~it incurs significant overhead due to a large number of VM exits, and (ii)~modern CVMs implement mitigations that can significantly slow down single-stepping attacks~\cite{wilke2024tdxdown,rauscher2025tdxploit}.

We therefore propose chained page arming, a single-step-free approach tailored to monitoring accesses to down projection weights from multiple vCPUs during multi-threaded, model-parallel LLM inference.
\Cref{fig:zapping} depicts the chain, consisting of five types of CVM-private pages:
\circledr{1}~an up projection weight page of $T^{(1)}$,
\circledr{2}~a code page that contains the definition of {\bf\small memset} invoked by each thread to zero an intermediate buffer during $D^{(1)}$,
\circledr{3}~a code page that contains the definition of the sparse down projection function executed by each thread,
\circledr{4}~the selected down projection weight pages of $T^{(1)}$ being monitored,
and \circledr{5}~an up projection weight page of $T^{(2)}$.

To monitor accesses to these pages, \coolname{} first identifies their GPAs using the method described in \Cref{sec:tracing:gpa}.
\coolname{} then begins monitoring LLM inference on prompt $(x_1, \dots, x_n)$ by arming \circledr{1}.
When a page fault occurs on \circledr{1}, \coolname{} disarms it and arms \circledr{2} and \circledr{5}.
This continues along the chain, with each fault disarming the faulting page and arming the next page(s) in the chain.
The disarmed pages must be armed again to monitor page accesses created by tokens subsequently processed.
To this end, we introduce two backward edges:
(i)~a fault triggered at \circledr{3} arms \circledr{2}, which allows \coolname{} to monitor page accesses created by the next prompt token processed in the batched down projection loop, $D^{(1)}(x_1), \dots, D^{(1)}(x_n)$,
and (ii)~a page fault at \circledr{5} arms \circledr{1}, allowing \coolname{} to monitor page accesses created by the next response token in the autoregressive generation loop, $T^{(1)}(x_n), \dots, T^{(1)}(x_{n+T-1})$.

The two code pages \circledr{2} and \circledr{3} executed by all threads serve as double-turnstile barriers~\cite{downey2008littlebookofsemaphores} during model-parallel execution of $D^{(1)}(x_1, \dots, x_n)$.
The page faults at these pages demarcate token boundaries in each vCPU, and the barriers force all vCPUs to synchronize before proceeding to the next prompt token.
This enables \coolname{} to aggregate partial access patterns of $D^{(1)}(x_i)$ from all vCPUs for each prompt token $x_i$.
Even when a vCPU does not access any monitored down projection weights for $x_i$, the barrier-induced page faults still mark when $D^{(1)}(x_i)$ begins and ends on that vCPU.

\section{Inverting Neuron Activation Traces}
\label{sec:inversion}
\noindent
\coolname{} then proceeds to this offline phase, where the neuron activation traces collected in the earlier online phase are inverted to reconstruct the original tokens.

\begin{algorithm}[t]
\footnotesize
\small
\caption{\coolname{'s} activation trace inversion.}
\label{alg:inversion:greedy}
\begin{algorithmic}[1]
\State \textbf{Input:} Activation traces $\{\tilde{B}^{(1)}(x_i)\}_{i=1}^{n+T-1}$, victim LLM, vocabulary $\mathcal{V}$
\State \textbf{Output:} Reconstructed token sequence $(\hat{x}_1, \dots, \hat{x}_{n+T-1})$
\For{$i = 1$ \textbf{to} $n+T-1$} \Comment{Loop for autoregressive reconstruction}
    \If{$i = 1$}
        \State $\mathbf{P}_i \gets \text{Uniform}(\mathcal{V})$ \Comment{No prior context}
    \Else
        \State $\mathbf{P}_i \gets \{P(v \mid \hat{x}_1, \dots, \hat{x}_{i-1}) : v \in \mathcal{V}\}$ \Comment{Full forward pass}
    \EndIf
    \State $(v_1, v_2, \dots, v_{|\mathcal{V}|}) \gets \text{Sort } \mathcal{V} \text{ by } \mathbf{P}_i \text{ (descending)}$
    \For{$j = 1$ \textbf{to} $|\mathcal{V}|$}
        \State $B_{v_j} \gets \tilde{B}^{(1)}(v_j)$ \Comment{First-layer forward pass}
        \If{$B_{v_j} = \tilde{B}^{(1)}(x_i)$}
            \State $\hat{x}_i \gets v_j$
            \State \textbf{break} \Comment{Stop at the first match}
        \EndIf
    \EndFor
\EndFor
\State \Return $(\hat{x}_1, \dots, \hat{x}_{n+T-1})$
\end{algorithmic}
\end{algorithm}

\subsection{Autoregressive Trace Inversion}
\label{sec:inversion:autoregressive}
\noindent We reconstruct the sequence $(\hat{x}_1, \dots, \hat{x}_{n+T-1})$ autoregressively starting from the first token $\hat{x}_1$.
Specifically, we invert ${\tilde{B}^{(1)}}(x_1)$ to reconstruct $\hat{x}_1$, $\tilde{B}^{(1)}(x_2)$ to reconstruct $\hat{x}_2$, and so on.
At iteration $i$, we condition the reconstruction of $\hat{x}_i$ on all previously reconstructed tokens $(\hat{x}_1, \dots, \hat{x}_{i-1})$.
We adopt this greedy, autoregressive approach, because the alternative, which treats all tokens as simultaneous variables and searches for a global optimum that elicits the same neuron activations, is computationally infeasible.
With the large vocabularies of modern LLMs, the time complexity of such a global search grows exponentially with the length of the token sequence.

A possible concern with the greedy approach is that an error in reconstructing one token could propagate and degrade the accuracy of subsequent token reconstruction.
In practice, however, we observe that such error propagation is limited (see \Cref{sec:evaluation:inversion}).
\coolname{'s} reconstruction relies on activations from the first layer of the model, which preserve direct information about the input token before extensive contextualization occurs in deeper layers\cite{he2025lawnexttoken}.

\subsection{Search-Based Trace Inversion}
\label{sec:inversion:search}

\noindent
At $i$th iteration of the autoregressive sequence reconstruction, \coolname{} searches for the token whose activation matches the observed activation trace $\tilde{B}^{(1)}(x_i)$.

\smallskip
\noindent\textbf{A Strawman Approach: Exhaustive Vocabulary Search.}
A straightforward approach is to compare $\tilde{B}^{(1)}(x_i)$ with the activation patterns of all possible tokens in the vocabulary.
That is, it performs a forward pass of the first layer of the victim LLM for every token in the vocabulary $\mathcal{V}$.
This yields $|\mathcal{V}|$ activation traces, which we then binarize and prune to obtain $\{\tilde{B}^{(1)}(v) \mid v \in \mathcal{V} \}$.
After this partial forward pass for each $v$, we compare its activation trace $\tilde{B}^{(1)}(v)$ with the observed activation trace $\tilde{B}^{(1)}(x_i)$.
If they match, $v$ is flagged as a candidate for successful reconstruction.
This approach suffers from two limitations, however: (i)~multiple candidate tokens may produce identical binary activation patterns, rendering the search ambiguous, and (ii)~the time complexity is prohibitively high for LLMs with a large $|\mathcal{V}|$.

\smallskip
\noindent\textbf{Our Approach: Guided Vocabulary Search.}
To resolve activation ambiguities, we exploit the target LLM's inherent next-token probability distribution over its vocabulary.
This distribution, calculated for the $i$th position conditioned on the previously recovered token sequence $(\hat{x}_1, \dots, \hat{x}_{i-1})$, naturally ranks all possible tokens by their contextual likelihood.
We examine candidates in descending probability order, terminating when a candidate's activation pattern matches the observed trace.
As we show in \Cref{sec:evaluation}, this prioritization achieves higher accuracy in the presence of pattern ambiguity, compared to an unguided, exhaustive search.

A potential drawback of this approach over the strawman one is that it additionally requires a full forward pass when reconstructing each token.
However, we observe that matching activations are often found within a few high-probability next tokens produced by the full forward pass.
This allows the search to terminate earlier than the naive method, effectively limiting the impact on inversion speed (see \Cref{sec:evaluation:ablation}).

\smallskip
\noindent\textbf{Alternative: Optimization-Based Trace Inversion.}
We also considered an optimization-based approach, with the optimization goal of minimizing the disparity between the generated activation and the target activation trace.
The inversion can be solved by initializing the embedding vector randomly, computing a loss based on the difference between the generated and target traces, and then backpropagating that loss to update the initialized embedding vector.
The final updated embedding vector after optimization would closely approximate the original token's embedding.
However, we ultimately rejected this approach because the optimization process yielded a weak loss signal.
Since the target activation trace is only available in its binarized form in our setting (i.e., a sequence of 0s and 1s, rather than continuous real values), the resulting objective function was too sparse to reliably guide the gradient-based optimization toward the correct embedding.

\section{Implementation Details}
\noindent
We implemented \coolname{} atop Linux Kernel Virtual Machine (KVM)~\cite{kvm} (v6.8.0) and QEMU~\cite{bellard2005qemu} (v8.2.50).

\smallskip
\noindent\textbf{Identifying Logical Block Addresses.}
To monitor the CVM's accesses to blocks that are loaded into the guest OS page cache and subsequently monitored by \coolname{} (i.e., the nodes in the page arming chain),
we compute the logical block address (LBA) of each target block as follows.
First, we read the guest disk image's partition table, identify the relevant partition, and retrieve the file's inode.
From this inode, we derive the file's block offset, and add the partition's starting LBA to obtain the file's starting LBA.
We then convert the byte offset of the target block within the file into a block offset, and add the file's starting LBA to obtain the target block's LBA.

\smallskip
\noindent\textbf{Finding GPAs of Loaded Blocks.}
We modified QEMU's {\it virtio\=/blk} device to intercept the CVM's accesses to target blocks.
Upon interception, it invokes \coolname{'s} kernel module to request arming of the bounce buffer page, and the last $k$ private pages below the high watermark upon bounce buffer faults.
The module exposes an {\it ioctl} interface for arming pages and retrieving the faulting page's address, allowing
\coolname{} to learn the private GPA of each loaded block.

\smallskip
\noindent\textbf{Chained Page Arming.}
Once the private GPAs of all nodes in the chain are known, \coolname{} activates the chain by sending these GPAs to its kernel module.
To efficiently track varying numbers of monitored down projection weight pages, \coolname{} maintains a per-CVM red-black tree of their GPAs.
Chained page arming is implemented primarily within the kernel module, by hooking the page fault handler to disarm the faulting page and arm the next page(s) in the chain.
We track the armed state of each monitored page, and use a per-page lock to ensure that only a single vCPU arms or disarms a page, even when multiple vCPUs fault on it concurrently.
The handler exits to the user-mode VMM (i.e., QEMU) where \coolname{} logs the accessed down projection weights and token changes.
These logs are then used in the offline phase and inverted to reconstruct the tokens.

\section{Evaluation}
\label{sec:evaluation}

\begin{table}
\caption{LLMs targeted in our evaluation.}
\label{tab:evaluation:models}
\centering
\footnotesize
\setlength\tabcolsep{0.1em}
\begin{threeparttable}
\begin{tabular}{|l|c|r|r|r|r|r|}

\hline
\makecell[cc]{Model} &
\makecell[cc]{\textls[-50]{Activation}} &
\makecell[cc]{\textls[-50]{Layers}\\($N$)} &
\makecell[cc]{\textls[-70]{Hidden Dim}\\($d_{\textnormal{model}}$)} &
\makecell[cc]{\textls[-50]{FFN Dim}\\($d_{\textnormal{ff}}$)} & 
\makecell[cc]{\textls[-50]{Neuron}\\\textls[-50]{Size}\textsuperscript{*}} &
\makecell[cc]{\textls[-50]{Vocab Size}\\($|\mathcal{V}|$)}\\
\hline
\sffamily\scriptsize \opt{-6.7b} & ReLU & 32 & 4,096 & 16,384 & 8,192 & 50,272 \\
\hline
\sffamily\scriptsize \relullama{-7B} & ReLU\textsuperscript{\dag} & 32 & 4,096 & 11,008 & 8,192 & 32,000 \\
\hline
\sffamily\scriptsize \textls[-50]{\nemotron{-3-8B-Base-4k}} & ReLU$^2$ & 32 & 4,096 & 16,384 & 8,192 & 256,000 \\
\hline
\sffamily\scriptsize \llama{-2-7b} & \textls[-50]{SwiGLU}\textsuperscript{\ddag} & 32 & 4,096 & 11,008 & 8,192 & 32,000 \\
\hline
\sffamily\scriptsize \gemma{-7b} & \textls[-50]{GeGLU}\textsuperscript{\ddag} & 28 & 3,072 & 24,576 & 6,144 & 256,000 \\
\hline

\end{tabular}
\end{threeparttable}
\scriptsize
\begin{tablenotes}[flushleft]
\item \textsuperscript{*} The size of each neuron is measured in bytes, and calculated as $d_{\textnormal{model}} \times 2$ bytes.
\item \textsuperscript{\dag} This model was ReLU-fied, after training.
\item \textsuperscript{\ddag} We applied magnitude-based thresholding.
\end{tablenotes}

\end{table}

\begin{figure*}[t]
  \centering
  \includegraphics[width=\textwidth]{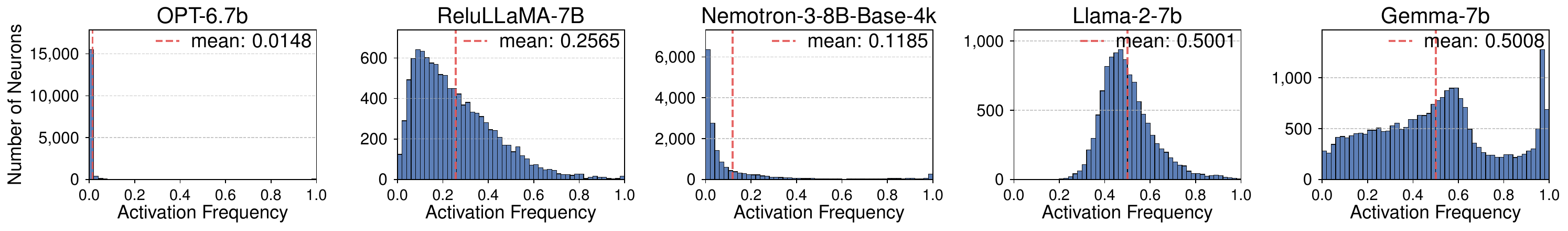}
  \caption{Distribution of activation frequencies (fraction of tokens activating each neuron) on the Wikipedia dataset. Red dashed lines indicate average frequencies.}
  \label{fig:evaluation:actfeq}
\end{figure*}

\smallskip
\noindent\textbf{Target Models.}
\label{sec:evaluation:targetmodels}
We used \smallopt{-6.7b}~\cite{zhang2022opt}, \smallrelullama{-7B}~\cite{sparsellm2023sparsellm}, \smallnemotron{-3-8B-Base-4k}~\cite{nvidia2023nemotron}, \smallllama{-2-7b}~\cite{touvron2023llama2openfoundation}, and \smallgemma{-7b}~\cite{google2024gemma} for our evaluation, with varying properties, as summarized in \Cref{tab:evaluation:models}.
Notably, these models use different activation functions: ReLU, ReLU$^2$, SwiGLU, and GeGLU.
For \smallllama{-2-7b} and \smallgemma{-7b} that use SwiGLU and GeGLU respectively, we applied magnitude-based thresholding that prunes the lowest 50\% of activations by magnitude in each layer, following Federici et al.~\cite{federici2025dip}.
The models have different numbers of neurons in each layer and different vocabulary sizes, while the size of each neuron is 8KiB in most models and 6KiB in \smallgemma{-7b}.
The models also exhibit different activation sparsity levels and patterns, per our own profiling shown in \Cref{fig:evaluation:actfeq}.
ReLU-variant models show skewed distribution with varying sparsity levels, whereas \smallllama{-2-7b} and \smallgemma{-7b} after pruning show much denser activation patterns, with mean frequencies near 0.5.

\smallskip
\noindent\textbf{Datasets.}
We used five datasets: three real-world datasets---Skytrax Reviews~\cite{danisman2019skytraxairlinereviews}, Medical WiKiDoc~\cite{han2023medalpaca}, and ECHR Law~\cite{chalkidis2019neurallegaljudgmentprediction}---and two synthetic datasets---Private Prompts~\cite{morris2023privateprompts} and System Prompt Leakage~\cite{chua2024systempromptleakage}.
These datasets contain sensitive prompts that elicit sensitive responses, and have therefore been used extensively in prior token leakage attacks~\cite{zhang2024output2prompt,morris2024lminversion,song2025earlybird,das2025systempromptextraction}.
Unless stated otherwise, we randomly sampled 100 prompts from each dataset for evaluation.
To train the LoRA adapters, we used the LaMP-4~\cite{salemi2024lamp}, a dataset widely used in prior work on personalized generation~\cite{tan2024democratizing,chen2024persona,xu2025personalized}.

\begin{figure*}[t]
  \label{fig:eval:speed}
  \centering
  \includegraphics[width=\textwidth]{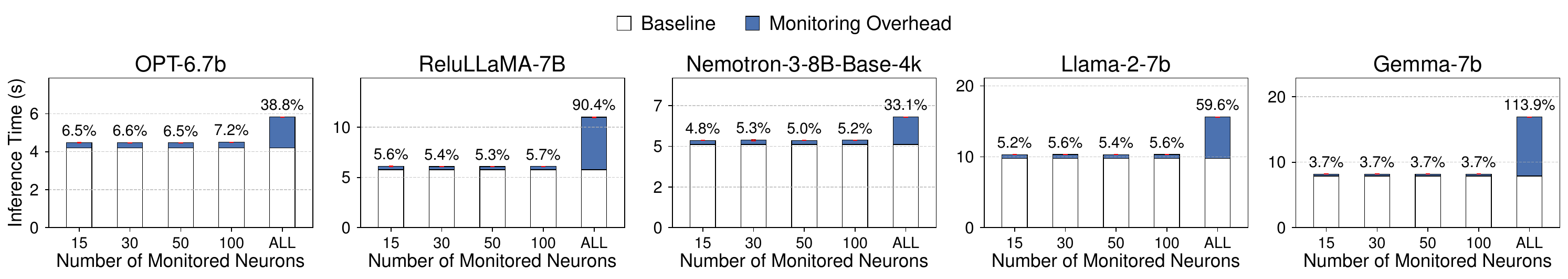}
  \caption{Wall-clock inference time before \& after \coolname{'s} monitoring. Results were obtained while each LLM was processing a prompt in Skytrax Reviews, averaged over 100 trials. The ``ALL'' column shows the overhead without selective monitoring.}
  \label{fig:evaluation:runtimeoverhead}%
\end{figure*}

\subsection{Inference-Time Overhead}
\noindent
We first quantify the inference-time overhead introduced by \coolname{'s} neuron activation monitoring during the online phase, i.e., during the victim LLM service's inference.

\smallskip
\noindent\textbf{Experimental Setup.}
We targeted PowerInfer~\cite{song2024powerinfer}, an optimized LLM serving system built on {\tt llama.cpp}~\cite{llama.cpp} that uses sparse down projection based on activation sparsity.
We incorporated magnitude-based thresholding into PowerInfer to induce activation sparsity for SwiGLU and GeGLU models.
While PowerInfer also supports sparse up projection using an activation predictor, we disabled this feature, because (i)~it degrades inference accuracy and (ii)~predictors are not available for certain models that we used in our evaluation.
We deployed our modified PowerInfer on a server equipped with dual Intel Xeon Gold 6548Y+ processors (128 logical cores in total) and 1 TiB of RAM.
We configured PowerInfer to use 16 threads, and ran it inside a 16-vCPU, 32GB-RAM CVM so as to enable different forms of parallelism during inference.
The vCPUs were pinned to physical CPU cores to ensure stable performance measurements.
We used the default configuration of PowerInfer otherwise.
We used a single dataset---Skytrax Reviews---for measuring and analyzing inference-time overhead, since our goal here is to quantify the overhead introduced by \coolname{'s} neuron activation monitoring, which is largely independent of the input prompts.

\smallskip
\noindent\textbf{Wall-Clock Inference Time.}
We first measure the wall-clock inference time, i.e., the time each LLM takes to consume a single prompt and generate 30 tokens, with and without \coolname{'s} neuron activation monitoring.
\Cref{fig:evaluation:runtimeoverhead} presents our measurements.
The overhead ranges from 3.7\% to 7.2\% depending on the model, when monitoring 15 to 100 neurons.
When monitoring 100 neurons, which was sufficient for \coolname{} to achieve up to 100\% reconstruction accuracy across all five models (see \Cref{sec:evaluation:inversion}), the overhead remains at this level.
The overhead increases up to 33\%--114\% when monitoring all first-layer FFN neurons (denoted by ALL).
The results show that our selective neuron monitoring (\Cref{sec:tracing:optimizations}) effectively reduces inference-time overhead, improving stealthiness.

\newcommand{\perfchildarrow}[1]{%
  \hspace{#1em}
  \tikz[baseline=-0.5ex, x=1em, y=1em]{
    \draw[line width=0.5pt, ->] 
      (0, 0.5) -- (0, 0) -- (0.6, 0);
  }%
}

\begin{table}[t]
\caption{The number of per-token page faults averaged over Skytrax Reviews. The ALL column monitors every neuron in the first FFN layer and serves as the baseline; percentages in parentheses are relative to this baseline.}
\label{tab:evaluation:pagefaults}
\centering
\footnotesize
\setlength\tabcolsep{0.1em}
\begin{threeparttable}
\begin{tabular}{|l|r|r|r|r|r|}

\cline{2-6}
\multicolumn{1}{c|}{}
&
\multicolumn{5}{c|}{\makecell[cc]{Number of Monitored Neurons}}
\\
\hline
\makecell[cc]{Model} &
\makecell[cc]{15} & \makecell[cc]{30} & \makecell[cc]{50} & \makecell[cc]{100} &
\makecell[cc]{ALL} \\
\hline
\begin{filecontents*}{figures/data/pertoken-pagefaults-skytrax-hist.csv}
Model,15 Neurons,15 Neurons_err,30 Neurons,30 Neurons_err,50 Neurons,50 Neurons_err,100 Neurons,100 Neurons_err,All first layer Neurons,All first layer Neurons_err
OPT-6.7b,46.04,0.23,53.43,0.04,61.85,0.25,81.61,0.26,256.13,1.63
ReluLLaMA-7B,43.35,0.11,55.76,0.19,71.07,0.12,107.45,0.09,6848.38,11.09
Nemotron-3-8B-Base-4k,41.22,0.02,43.94,0.68,45.98,0.26,49.39,0.05,408.88,2.76
Llama-2-7b,41.72,0.30,49.12,0.89,58.18,0.70,82.33,1.27,5539.21,1.59
Gemma-7b,41.22,0.04,48.93,0.07,58.70,0.13,81.73,0.40,12320.26,3.23
\end{filecontents*}
\csvreader[
    head to column names,
    late after line=\\\hline
]{figures/data/pertoken-pagefaults-skytrax-hist.csv}{Model=\model,15 Neurons=\fifteen,15 Neurons_err=\fifteenerr,30 Neurons=\thirty,30 Neurons_err=\thirtyerr,50 Neurons=\fifty,50 Neurons_err=\fiftyerr,100 Neurons=\hundred,100 Neurons_err=\hundrederr,All first layer Neurons=\alln,All first layer Neurons_err=\allnerr}%
{\ifthenelse{\equal{\model}{OPT-6.7b}}{\sffamily\scriptsize \opt{-6.7b}}{}%
\ifthenelse{\equal{\model}{ReluLLaMA-7B}}{\sffamily\scriptsize \relullama{-7B}}{}%
\ifthenelse{\equal{\model}{Nemotron-3-8B-Base-4k}}{\sffamily\scriptsize \textls[-70]{\nemotron{-3-8B-Base-4k}}}{}%
\ifthenelse{\equal{\model}{Llama-2-7b}}{\sffamily\scriptsize \llama{-2-7b}}{}%
\ifthenelse{\equal{\model}{Gemma-7b}}{\sffamily\scriptsize \gemma{-7b}}{}%
 & \makecell[r]{\pgfmathprintnumber[fixed,fixed zerofill,precision=1]{\fifteen}\err{\pgfmathprintnumber[fixed,fixed zerofill,precision=1]{\fifteenerr}} \\ (\pgfmathparse{(\fifteen/\alln)*100}\pgfmathprintnumber[fixed,fixed zerofill,precision=2]{\pgfmathresult}\%)}
 & \makecell[r]{\pgfmathprintnumber[fixed,fixed zerofill,precision=1]{\thirty}\err{\pgfmathprintnumber[fixed,fixed zerofill,precision=1]{\thirtyerr}} \\ (\pgfmathparse{(\thirty/\alln)*100}\pgfmathprintnumber[fixed,fixed zerofill,precision=2]{\pgfmathresult}\%)}
 & \makecell[r]{\pgfmathprintnumber[fixed,fixed zerofill,precision=1]{\fifty}\err{\pgfmathprintnumber[fixed,fixed zerofill,precision=1]{\fiftyerr}} \\ (\pgfmathparse{(\fifty/\alln)*100}\pgfmathprintnumber[fixed,fixed zerofill,precision=2]{\pgfmathresult}\%)}
 & \makecell[r]{\pgfmathprintnumber[fixed,fixed zerofill,precision=1]{\hundred}\err{\pgfmathprintnumber[fixed,fixed zerofill,precision=1]{\hundrederr}} \\ (\pgfmathparse{(\hundred/\alln)*100}\pgfmathprintnumber[fixed,fixed zerofill,precision=2]{\pgfmathresult}\%)}
 & \makecell[r]{\pgfmathprintnumber[fixed,fixed zerofill,precision=1]{\alln}\err{\pgfmathprintnumber[fixed,fixed zerofill,precision=1]{\allnerr}}}}%
\end{tabular}
\scriptsize
\end{threeparttable}

\end{table}

\smallskip
\noindent\textbf{Number of Per-Token Page Faults.}
We now measure the number of page faults triggered during LLM inference while monitoring varying numbers of neurons.
We report per-token page fault counts, i.e., the number of page faults triggered while processing each token, because the same prompt can be tokenized into different lengths depending on the model (e.g., 64 tokens for \smallrelullama{} and \smallllama{}, vs. 56 tokens for \smallopt{}, \smallnemotron{}, and \smallgemma{}).
\Cref{tab:evaluation:pagefaults} presents the results.
As expected, the per-token page fault count rises as more neurons are monitored.
When monitoring all neurons in the first-layer FFN, the number of page faults increases substantially for \smallrelullama{-7B}, \smallllama{-2-7b}, and \smallgemma{-7b}, which are less sparse than the other two models (see \Cref{fig:evaluation:actfeq}).
These results demonstrate that our selective neuron monitoring substantially reduces page faults and thus inference-time overhead.

\begin{table}[t]
\caption{Breakdown of total CPU time during \smallopt{-6.7b}'s inference on a Skytrax Reviews prompt, shown for the runs with the minimum, median, and maximum total CPU time across 10 trials.}
\label{tab:evaluation:breakdown}
\centering
\footnotesize
\setlength\tabcolsep{0.07em} 
\begin{threeparttable}
\begin{tabular}{|l|rr|rr|rr|}
\hline
{Time spent by ...} &
\multicolumn{2}{c|}{{Min-time run}} &
\multicolumn{2}{c|}{{Median-time run}} &
\multicolumn{2}{c|}{{Max-time run}} \\
\hline

\begin{filecontents*}{figures/data/overhead-breakdown.csv}
Indent,Component,MinTime,MinPct,MedTime,MedPct,MaxTime,MaxPct,Line
0,Guest vCPU,69918,96.05,70368,95.59,70032,94.61,1
0,Host User-Mode,507,0.70,713,0.97,887,1.20,1
0,Host Kernel-Mode,2367,3.25,2535,3.44,3100,4.19,0
1,Barrier Wait,581,0.80,762,1.04,1028,1.39,0
1,IPI Calls,9,0.01,9,0.01,8,0.01,0
1,SEAMCALLs,27,0.04,26,0.04,26,0.03,0
1,Others,1751,2.41,1738,2.36,2038,2.75,1
0,Total CPU Time,72792,100.00,73615,100.00,74020,100.00,2
\end{filecontents*}
\csvreader[
    head to column names,
    late after line= 
]{figures/data/overhead-breakdown.csv}{
    Indent=\indentval,
    Component=\component,
    MinTime=\mintime, MinPct=\minpct,
    MedTime=\medtime, MedPct=\medpct,
    MaxTime=\maxtime, MaxPct=\maxpct,
    Line=\lineval
}%
{%
    \gdef\currRow{%
        \ifthenelse{\equal{\indentval}{1}}{\hspace{.5em}- }{}%
        \component &
        \pgfmathprintnumber[fixed, precision=0, fixed zerofill, 1000 sep={,}]{\mintime} &
        (\pgfmathprintnumber[fixed, precision=2, fixed zerofill]{\minpct}\%) &
        \pgfmathprintnumber[fixed, precision=0, fixed zerofill, 1000 sep={,}]{\medtime} &
        (\pgfmathprintnumber[fixed, precision=2, fixed zerofill]{\medpct}\%) &
        \pgfmathprintnumber[fixed, precision=0, fixed zerofill, 1000 sep={,}]{\maxtime} &
        (\pgfmathprintnumber[fixed, precision=2, fixed zerofill]{\maxpct}\%)%
    }%
    \currRow
    \ifthenelse{\equal{\lineval}{1}}{\\\hline}{%
        \ifthenelse{\equal{\lineval}{2}}{}{\\}%
    }%
}%
\\\hline

\end{tabular}
\end{threeparttable}
\end{table}

\smallskip
\noindent\textbf{Overhead Breakdown.}
We further analyze the sources of inference-time overhead, when \coolname{} monitors 100 neurons during \smallopt{-6.7b}'s inference with 16 vCPUs on a Skytrax Reviews prompt.
We measure per-core CPU times by instrumenting several kernel functions with Tracepoint~\cite{tracepoint}, and attaching eBPF~\cite{ebpf} programs to compute the time elapsed between them.
\Cref{tab:evaluation:breakdown} shows the breakdown of the total CPU time for three runs: those with the minimum, median, and maximum total CPU time.
The dominant off-VM overhead stems from barrier synchronization used to demarcate prompt tokens during batched prefill.
The contribution of IPI calls and SEAMCALLs is not significant: the overhead of IPI calls for TLB shootdowns and of arming and disarming pages is small compared to the total barrier synchronization overhead for long prompts.
However, as the number of monitored neurons increases, we expect their overhead to become more pronounced.
Finally, the impact on vCPU execution time due to cache pollution or TLB shootdowns appears negligible, likely due to the memory-bound nature of LLM inference.

\begin{figure*}[t]
	\centering
	\includegraphics[width=\textwidth]{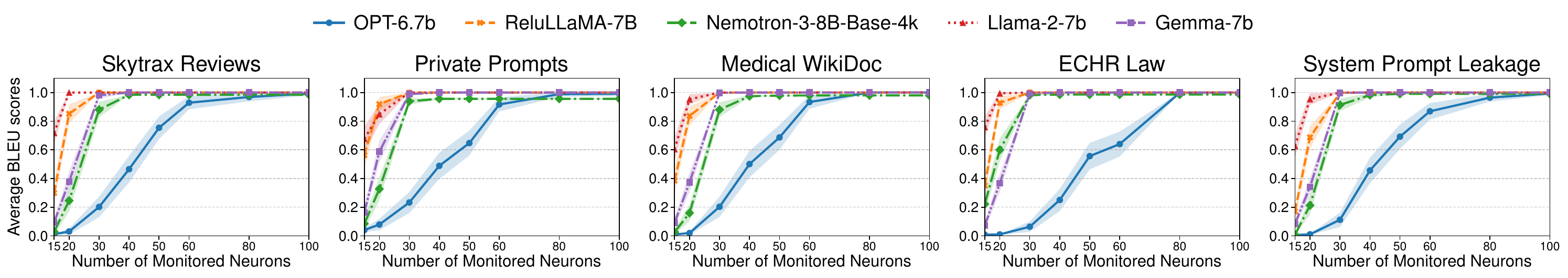}
	\caption{Reconstruction accuracy across five datasets, with varying numbers of monitored neurons. Measured by BLEU scores.}
	\label{fig:evaluation:bleualldataset}
\end{figure*}

\subsection{Reconstruction Accuracy}
\label{sec:evaluation:inversion}

\noindent We now evaluate the reconstruction accuracy of \coolname{'s} inversion during the offline phase.

\smallskip
\noindent\textbf{Experimental Setup.}
We inverted the collected traces using an NVIDIA H100 GPU with a 94GB VRAM.
The software environment was PyTorch 2.7.1 (built with CUDA 12.4) with NVIDIA driver 550.67.
To evaluate the reconstruction accuracy of our inversion methods, we use {BLEU}~\cite{papineni2002bleu},
which assesses the $n$-gram overlap between the original and reconstructed token sequences, effectively indicating the success rate of token recovery.
We used the Wikipedia dataset~\cite{wikidump} with approximately 135,000 tokens to compute joint entropy used in neuron selection (see \Cref{sec:tracing:optimizations}).
We begin with the top 15 neurons ranked by joint entropy, which represents the theoretical minimum required to uniquely identify all 32,000 ($\approx2^{15}$) tokens in the vocabulary.
This minimum requirement corresponds to the smallest vocabulary size among the models we evaluated, \smallrelullama{-7B} and \smallllama{2-7b}.
We then gradually increase the number of selected neurons up to 100.

\smallskip
\noindent\textbf{Quantitative Results.}
\Cref{fig:evaluation:bleualldataset} presents the reconstruction accuracy across all target models and datasets.
With only 100 monitored neurons, \coolname{} consistently achieves near-100\% reconstruction accuracy across all models and datasets that were evaluated.
Notably, \smallopt{-6.7b}, which exhibits the highest activation sparsity, requires 80 or more neurons to approach this level, whereas other models with lower sparsity achieve comparable accuracy with only 30 neurons.
This indicates that reconstructing highly sparse activations requires monitoring more neurons.
We also find that vocabulary size has limited influence on our search-based reconstruction.
Although the vocabulary of \smallnemotron{} and \smallgemma{-7b} is an order of magnitude larger than that of other models, their reconstruction performance remains comparable, suggesting that activation patterns across tokens are sufficiently distinct.

\newcommand{\mismatch}[1]{\textit{\textcolor{red}{\uline{#1}}}}
\newcommand{\orig}[1]{\textit{\uline{#1}}}
\newcommand{\response}[1]{\textbf{#1}}
\begin{table*}[t]
\caption{Examples of tokens reconstructed by inverting activation traces from 15 to 40 first-layer FFN neurons across five target LLMs. All prompts are from the Private Prompts dataset to illustrate the privacy risk.}
\label{tab:evaluation:reconstructionexamples}
\centering
\footnotesize
\setlength\tabcolsep{0.2em}
\begin{threeparttable}
\begin{tabular}{|c|c| @{\hspace{.5em}} m{0.402\textwidth} @{\hspace{.7em}} c @{\hspace{.7em}} m{0.402\textwidth}|}
\hline
\makecell[cc]{Target \\ Model} &
\makecell[cc]{\# of \\ Neurons} &
\makecell[cc]{{Original Prompt \& {\bf Response}}} & &
\makecell[cc]{{Reconstructed Prompt \& {\bf Response}}} \\
\hline
\scriptsize\opt{-6.7b} & 40 &
What is a question which results in the answer of ``19 April 1984''? Answer:\response{ The date of the first ever Test match between India and Pakistan.} & \large\MVRightarrow &
What is a question which results in the answer of ``19 April \mismatch{2001}''? Answer:\response{ The date of the first ever Test match between India and Pakistan.} \\
\hline
\makecell[cc]{\scriptsize\relullama{} \\ \scriptsize\textsf{-7B}} & 30 &
Please add punctuation to this: Latest visit November 2227\response{.\textbackslash{}n Latest visit: November 22, 2027.} & \large\MVRightarrow &
Please add punctuation to this: Latest visit \mismatch{September} 2227\response{.\textbackslash{}n Latest visit: November 22, 2027.} \\
\hline
\makecell[cc]{\scriptsize\nemotron{-3} \\ \scriptsize\textsf{-8B-Base-4k}} & 20 &
If Sarah Smith needs to buy 3 loaves of bread at the store, and each loaf costs \$2, how much money will Sally spend in total?\response{\textbackslash{}n Answer: 3 loaves of bread * \$2 each = \$6} & \large\MVRightarrow &
If Sarah Smith needs to buy 3 loaves of bread at the store, and each loaf costs \$2, how much money will Sally spend in total?\response{\textbackslash{}n Answer: 3 loaves of bread * \$2 each = \$\mismatch{amp}} \\
\hline
\scriptsize\llama{-2-7b} & 15 &
what was the number one song on 17 April 1970??\response{\textbackslash{}n The number one song on 17 April 1970 was ``The Candy Man'' by Sammy Davis Jr.} & \large\MVRightarrow &
what was the number one song on 17 April 1970??\response{\textbackslash{}n The number one song on 17 April 1970 was ``The Candy Man'' by Sammy Davis Jr.}
\\
\hline
\scriptsize\gemma{-7b} & 30 & Write an email with the following subject: Joint MSWG \& S\&PWG meeting, January\response{ 2011.\textbackslash{}n\textbackslash{}n The email should be sent to the following email addresses:\textbackslash{}n\textbackslash{}n * s\&pwg-chair@w3}
& \large\MVRightarrow & Write an email with the following subject: Joint MSWG \& S\&PWG meeting, January\response{ 2011.\textbackslash{}n\textbackslash{}n The email should be sent to the following email addresses:\textbackslash{}n\textbackslash{}n * s\&pwg-chair@w3}
\\
\hline
\end{tabular}
\end{threeparttable}
\end{table*}

\smallskip
\noindent\textbf{Qualitative Examples.}
We showcase several reconstructed prompts and responses in \Cref{tab:evaluation:reconstructionexamples}.
As shown, \coolname{} accurately reconstructs not only prompt tokens but also response tokens, and recovers both common filler words (e.g., ``a'', ``the'', ``is'') and highly specific tokens such as domain-specific terms and proper nouns (e.g., ``MSWG'').
Moreover, \coolname{} maintains high accuracy even for long sequences containing many tokens.
Finally, \coolname{} successfully reconstructs subsequent tokens even after a reconstruction error, validating the limited error propagation discussed in \Cref{sec:inversion:autoregressive}.

\begin{figure}[t]
\centering
\begin{minipage}{0.49\textwidth}
  \vspace{-3em}
  \centering
  \hspace{-0.2em}\includegraphics[width=0.995\textwidth]{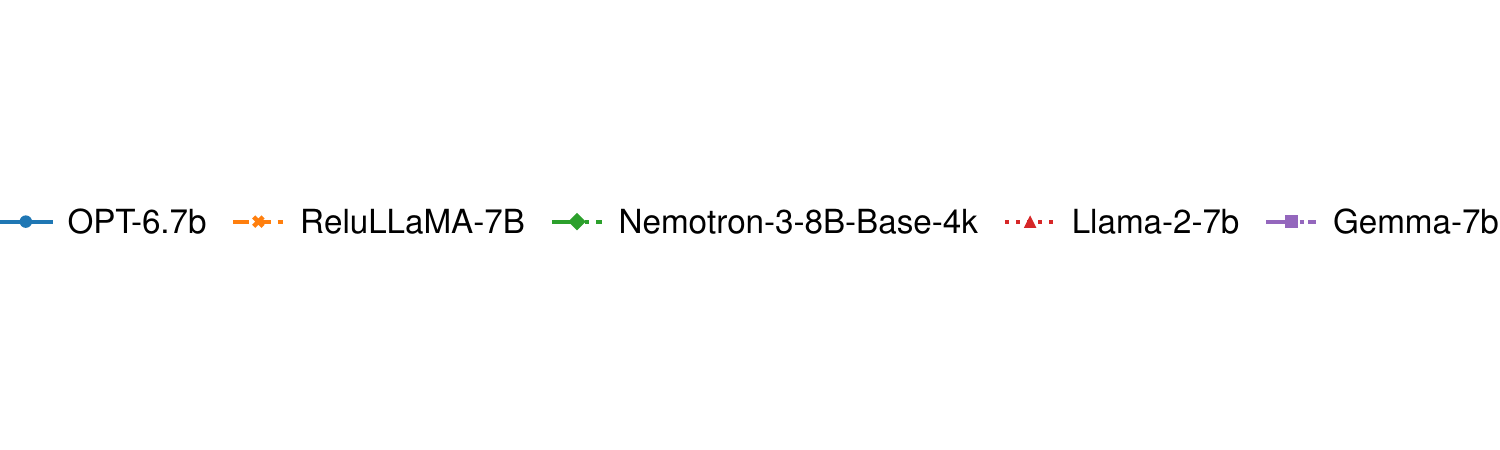}
  \vspace{-5.3em}
\end{minipage}
\subfloat[][After adapting models with private LoRA adapters.]%
{%
  \includegraphics[width=0.226\textwidth]{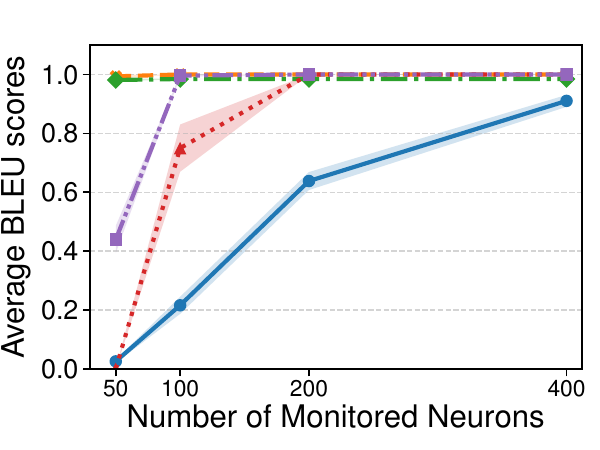}%
  \label{fig:evaluation:lorableu}%
}%
\hfill
\subfloat[][After offloading cold neuron computation.]%
{%
  \includegraphics[width=0.226\textwidth]{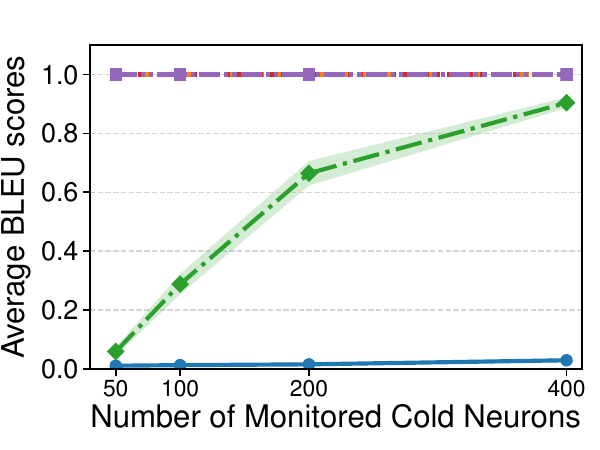}%
  \label{fig:evaluation:hotcoldbleu}
}
\caption{Reconstruction accuracy obtained after (a)~adapting the base models with private LoRA adapters with rank 8, and (b)~offloading cold neuron computation to the CPU, measured by BLEU scores on Skytrax Reviews.}
\label{fig:evaluation:loraandcoldBLEU}
\end{figure}

\subsection{Robustness to Fine-tuning \& Partial CPU Offloading}
\noindent
We evaluate our attack under two additional deployment scenarios that restrict the attacker's visibility: (i)~the base model is fine-tuned with private LoRA adapters unknown to the attacker, and (ii)~only infrequently activated (i.e., \emph{cold}) neurons are offloaded to the CPU, leaving the rest \emph{hot} neurons outside the attacker's observation scope.

\smallskip
\noindent\textbf{Experimental Setup.}
To evaluate the effectiveness of \coolname{} on LoRA-adapted models, we replicated the LoRA fine-tuning setup for personalized news headline generation from Tan et al.~\cite{tan2024democratizing}, in which rank-8 adapters are applied to the attention layers and trained for two epochs on the LaMP-4 dataset~\cite{salemi2024lamp}, with a learning rate of 1e-5 and a batch size of 8.
We confirmed successful fine-tuning by reproducing the final performance numbers reported in the original study~\cite{tan2024democratizing}, and examined \coolname{'s} token reconstruction performance without access to the adapters.

In the CPU offloading experiment, we first profiled neuron activations using the Wikipedia dataset~\cite{wikidump} to identify cold neurons.
Assuming that GPU memory can accommodate half of the model parameters, we offloaded the 50\% least frequently activated neurons in each layer to the CPU.
We then recomputed the joint entropy over these offloaded cold neurons on the same Wikipedia dataset to select which ones to monitor.

\smallskip
\noindent\textbf{Reconstruction Accuracy with LoRA Fine-tuning.}
We collected neuron activation traces of each model fine-tuned with a private LoRA adapter through \coolname{'s} neuron activation oracle, and then reconstructed the tokens that produced these activations with its base model through our search-based, autoregressive inversion~(see \Cref{sec:inversion}).
As shown in \Cref{fig:evaluation:lorableu}, BLEU scores improve as more first-layer neurons are monitored, reaching up to 1.0000.
All models except \smallopt{-6.7b} reach a BLEU score of 0.9845 or higher with 200 monitored neurons; \smallopt{-6.7b} reaches 0.9103 with 400 neurons.
This robustness suggests that fine-tuning typically alters only a limited subset of model parameters, largely preserving (i)~first-layer activation patterns, and (ii)~next-token probability rankings.

\smallskip
\noindent\textbf{Reconstruction Accuracy with Partial CPU Offloading.}
\Cref{fig:evaluation:hotcoldbleu} shows reconstruction accuracy when the attacker observes cold neuron activations.
As expected, this setup is more challenging than monitoring all neurons.
Even with 400 monitored neurons, \smallnemotron{-3-8B-Base-4k} reaches a BLEU score of 0.9040, and \smallopt{-6.7b} performs significantly worse, reaching only 0.0296.
This means that, in these models, neurons with high joint entropy (i.e., discriminative w.r.t. the input tokens) coincide with hot neurons, which are no longer observable to the attacker.
For the rest of three models, however, cold neurons are sufficiently discriminative, reaching a BLEU score of 1 with only 50 neurons.

\subsection{Ablation Studies}
\label{sec:evaluation:ablation}

\begin{figure}[t]
\centering
\begin{minipage}{0.49\textwidth}
  \vspace{-3em}
  \centering
  \hspace{-0.2em}\includegraphics[width=0.995\textwidth]{figures/varying-neurons-bleu-legend.pdf}
  \vspace{-5.3em}
\end{minipage}
\subfloat[][Random neuron selection (baseline).]%
{%
  \includegraphics[width=0.226\textwidth]{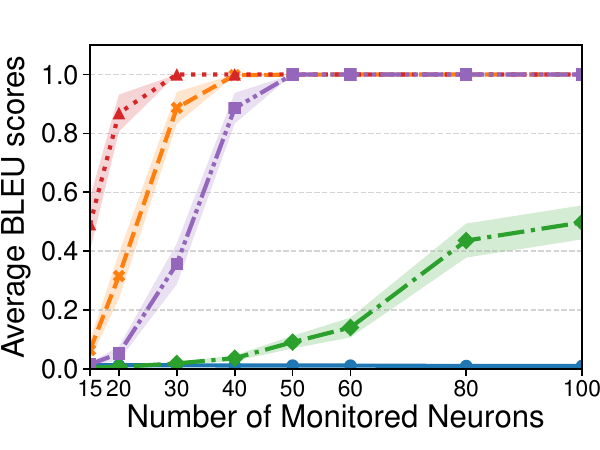}%
  \label{fig:evaluation:jointvaryingneuronsBLEU}%
}%
\hfill
\subfloat[][Entropy-based neuron selection (ours).]%
{%
  \includegraphics[width=0.226\textwidth]{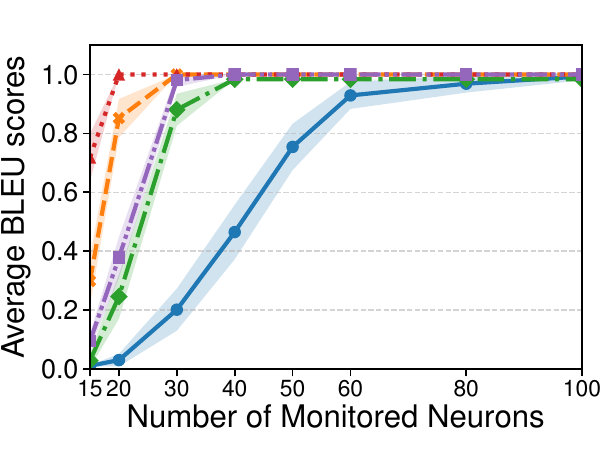}%
  \label{fig:evaluation:randomvaryingneuronsBLEU}%
}
\caption{Reconstruction accuracy obtained via (a)~random neuron selection and (b)~\coolname{'s} entropy-based neuron selection. Measured by BLEU scores on Skytrax Reviews.}
\label{fig:evaluation:randomcomparisonBLEU}
\end{figure}

\noindent\textbf{Random vs. Entropy-based Neuron Selection.}
\noindent
We evaluate our entropy-based neuron selection technique by comparing it against random selection and analyzing how reconstruction accuracy differs between the two.
\Cref{fig:evaluation:randomcomparisonBLEU} depicts their reconstruction accuracy measured in BLEU scores.
All models show improved performance over random selection, particularly \smallopt{-6.7b} and \smallnemotron{-3-8B-Base-4k}.
For \smallllama{-2-7b}, only 20 neurons already yield a BLEU score of 1.0, and all models except \smallopt{-6.7b} achieve satisfactory reconstruction with as few as 30 neurons.
The \smallopt{-6.7b} model requires more neurons, likely due to its highly sparse activations.
Nevertheless, 100 neurons constitute only 0.61\% of the 16,384 neurons in the first layer of \smallopt{-6.7b}.
Considering that random selection for \smallopt{-6.7b} did not achieve a BLEU score of 0.1 even with 100 neurons, our entropy-based selection yields a substantial improvement in reconstruction performance.

\begin{figure}[t]
\centering
\begin{minipage}{0.49\textwidth}
  \centering
  \includegraphics[width=0.75\textwidth]{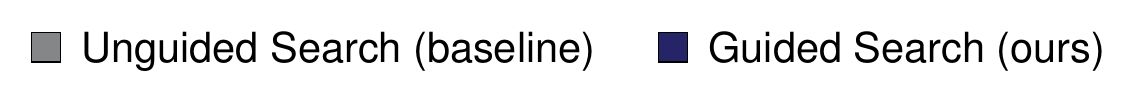}
  \vspace{-2.5em}
\end{minipage}
\subfloat[][Reconstruction accuracy.]%
{%
  \includegraphics[width=0.236\textwidth]{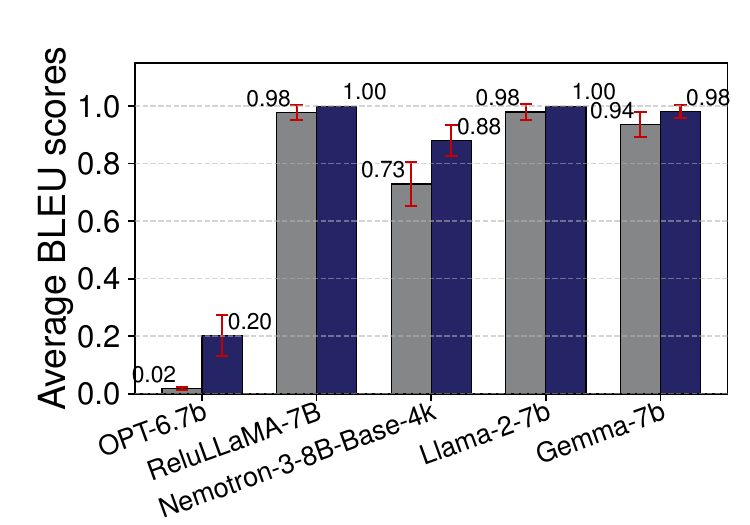}%
  \label{fig:evaluation:orderingBLEU}%
}%
\hfill
\subfloat[][Wall-clock inversion time.]%
{%
  \includegraphics[width=0.236\textwidth]{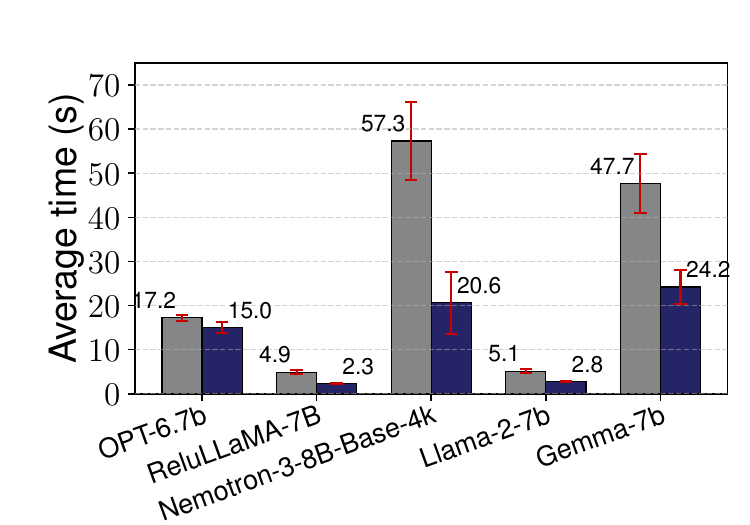}%
  \label{fig:evaluation:orderingtimes}%
}
\caption{Effectiveness of \coolname{'s} guided search in terms of (a)~reconstruction accuracy and (b)~computation time. Results were obtained by inverting binary activation traces of top-30 high-entropy neurons on Skytrax Reviews.}
\label{fig:evaluation:guidedsearch}
\end{figure}

\smallskip
\noindent\textbf{Unguided vs. Guided Search.}
Search-based activation inversion suffers from multiple matches: multiple tokens can produce identical activation pattern once binarized.
This ambiguity increases when only a subset of neurons is monitored.
\coolname{} addresses this by prioritizing candidate tokens according to their logits (see \Cref{sec:inversion:search}).
To validate this strategy, we conducted experiments that compare the BLEU scores obtained when monitoring 30 neurons, using either the vocabulary's token ID order (denoted by {\sf\small Unguided Search}) or our logit-based ordering ({\sf\small Guided Search}).
The results presented in \Cref{fig:evaluation:orderingBLEU} demonstrate that our guided search consistently improves reconstruction accuracy.

We also evaluate how quickly activation traces extracted from the victim can be inverted.
\Cref{fig:evaluation:orderingtimes} compares the wall-clock time required to invert activation traces using unguided and guided search.
Although guided search requires a full forward pass to compute next-token probabilities, it achieves faster inversion by terminating the search earlier.
Since inversion can be performed offline, our attack's success is largely insensitive to inversion speed.
We nevertheless report these results (i)~to demonstrate that inversion can be completed within a practical timeframe, and (ii)~to support future research on real-time or dynamically-adapting inversion attacks, whose covert operation may benefit from faster inversion.

\section{Discussion \& Limitations}
\label{sec:discussion}

\noindent\textbf{Potential Mitigations.}
\noindent
A straightforward mitigation against our attack is to disable sparsity-exploiting optimizations altogether in LLM serving systems.
However, doing so forfeits their substantial efficiency gains (e.g., up to $2\times$ inference speedup~\cite{liu2023dejavu}).

A more lightweight alternative is to randomize the GPAs of neurons, preventing the attacker from inferring neuron activations through page-access monitoring.
This can be achieved by randomizing the placement of the down projection weights within the model file, and keeping the file (or the entire block device) encrypted at rest on the host side.
Block encryption alone, e.g., via \emph{dm-crypt}~\cite{dmcrypt67}, is insufficient, since it preserves the block-level layout of the model.
Such randomization is also static---the layout does not change across requests---so an attacker with query access to the victim (which we do not assume) can still profile neuron activation patterns using chosen prompts.
A more principled, albeit more expensive, defense is to dynamically reshuffle access patterns via oblivious RAM techniques~\cite{goldreich1996oram,rane2015raccoon,nayak2017hop}.

One might also defeat our attack by executing LLM inference entirely on GPUs.
However, when a model exhibits high sparsity (e.g., fewer than 10\% of neurons active, as in OPT models), it is often more efficient to perform sparse matrix-vector multiplication on the CPU.
Indeed, many LLM-serving systems already offload sparse computation
to the CPU to better saturate CPU/GPU compute and reduce GPU I/O bandwidth and memory demands~\cite{xue2024powerinfer2,sheng2023flexgen,lee2024infinigen,park2024improving}.
Our evaluation shows that even when only 50\% of the FFN computation runs on the CPU, our attack still achieves high reconstruction accuracy across most models.
Finally, recent work by Chrapek et al. shows that even fully CPU-based LLM inference inside CPU TEEs can be more cost-efficient than GPU TEEs such as those of NVIDIA's H100~\cite{chrapek2025confidentialllminference}.

Another mitigation is to keep the model weights private, since our attack assumes the adversary has weight access.
However, this defense can often be circumvented.
If the victim uses an open-source model but the attacker lacks knowledge of the model, our attack can be combined with prior techniques that infer model hyperparameters and architecture~\cite{hua2018reversecnn,yan2020cachetelepathy,gao2024deeptheft,liu2024deepcache},
enabling the attacker to fingerprint and retrieve the corresponding model from a public repository.
For proprietary models, previous work on black-box intermediate activation reconstruction attacks~\cite{dong2025depth} can be adopted.

\smallskip
\noindent\textbf{Other Side-Channel Attack Vectors.}
\label{sec:discussion:othersidechannels}
\noindent
Controlled channels are not the only means through which neuron activation patterns may be leaked.
Hardware-based off-chip memory access side channels~\cite{lee2020offchip,maas2013phantom} (or even physical bus interposition~\cite{chuang2026teefail}), as demonstrated by Hua et al. in model stealing attacks~\cite{hua2018reversecnn}, can also expose neuron activation patterns.
CPU cache side channels~\cite{yarom2014flushreload,gruss2016flushflush} may also reveal neuron access patterns, e.g., by flushing page table entries from CPU caches~\cite{van2017withoutpagefaults}.
Recently, researchers also found that Intel~TDX gives adversaries a new means to invalidate a cache line~\cite{intel-tdx-google-review}, which could enhance flush-based attacks~\cite{yarom2014flushreload,gruss2016flushflush}.
Concurrent work by Hornetz et al. combines controlled channels with CPU cache side channels to recover prompts by monitoring memory accesses during tokenization of prompts~\cite{hornetz2026tdxray}.
Spielman et al.~\cite{spielman2025controlledmodelstealing}, on the other hand, combine controlled channels with the single-stepping primitive against enclaves to recover model parameters.

Ciphertext side channels and floating-point timing side channels
could be another potential vector for leaking activation patterns.
Prior work used the former to leak DNN intelligence~\cite{yuan2024hypertheft}, and the latter to reverse-engineer DNNs~\cite{gongye2020floatingpoint}.
Yan et al. combined ciphertext side channels with the attacker's ability to relocate pages to infer partial information about a victim's prompt~\cite{yan2025relocatevote}.
Unlike our approach, their attack exploits leakage from sparse activations, rather than selective neuron accesses during sparse down projection.
Further investigation of these and other side channel attack vectors remains an open direction for future research.

\smallskip
\noindent\textbf{Other Risky Inference-Time Optimizations.}
\label{sec:discussion:otheroptimizations}
\noindent
Our work reveals a fundamental tension between inference-time optimizations increasingly used in LLM serving systems and their implications on input or model privacy.
While we focused on sparse matrix-vector computation, any optimization that introduces input- or model-dependent behavior can potentially be exploited to leak information about user prompts or models.

Caching mechanisms such as KV cache, for example, introduce input-dependent behavior.
A line of work showed that such behavior are exposed via timing side channels, which can in turn be exploited to recover input prompts~\cite{song2025earlybird,zheng2024inputsnatch,wu2025promptpeek,luo2026kvinversion}.
Mixture-of-Experts (MoE) models (including MoE-fied ones~\cite{zhang2022moefication,szatkowski2024d2dmoefication}) also exhibit input-dependent behavior during expert selection.
Prior work has shown that per-batch routing in MoE models can be abused to drop~\cite{hayes2024bufferovfmoe} and even steal~\cite{yona2024promptstealingfrommoe,ding2025moecho} tokens from other co-tenants scheduled in the same batch as the attacker.
By abusing speculative decoding~\cite{leviathan2023specdecode}, Wei~et~al. showed (i)~that side channels can be used to distinguish successful and unsuccessful speculation, and (ii)~that information about prompts can be leaked~\cite{wei2024specdecodingidechannel}.
Hua~et~al. showed that, by abusing the zero-pruning optimization implemented in accelerators, the ratio between the weight and bias parameters can be revealed through side channels~\cite{hua2018reversecnn}.
These works, including ours, highlight the need to reevaluate the security of LLM serving systems with modern optimizations.

\section{Related Work}

\subsection{Prompt Extraction Attacks}
\noindent
Our work relates to a broader line of work on prompt extraction attacks against LLMs, among which we highlight black-box, API-only attacks not covered in \Cref{sec:overview}.
These attacks target system prompts, i.e., prompts that are prepended to user inputs by the LLM serving system, which are often considered proprietary and secret.
Researchers proposed to extract system prompts via adversarial or jailbreak-style queries~\cite{zhang2024systempromptextraction,hui2024pleak}, by observing the model's output probabilities~\cite{morris2024lminversion} or output tokens~\cite{zhang2024output2prompt}.
Unlike these attacks, ours requires no access to the LLM service interface, and can recover both system and user prompts, as well as the model's response, by monitoring the victim's memory accesses throughout prefill and decoding.

\subsection{Confidential Machine Learning}
\label{sec:relatedwork:conf}
\noindent
Our work relates to a rich body of work that explores deploying DNN serving systems inside TEEs to protect models and user inputs~\cite{mo2024mlwithcc}.
A line of work focuses on on-device DNN inference for edge or mobile platforms~\cite{mo2020darknetz,sun2022shadownet,siby2024guarantee,zhang2024teeslice,shen2022soter,moon2025asgard,wang2026tzllm}.
Another line of work targets cloud scenarios, using Intel~SGX~\cite{lee2019occlumency} and, more recently, GPU TEEs~\cite{tan2025pipellm} and NPU TEEs~\cite{feng2024snpu}, for protecting models or user data during DNN computation.
Notably, NVIDIA has introduced confidential computing capabilities in their H100 GPUs~\cite{nvidia-h100}.
Tan et al. identified inefficiencies in encrypted memory transfers between CPU TEEs and H100 GPU TEEs, and proposed pipelining techniques to accelerate LLM inference on CPU-GPU TEEs~\cite{tan2025pipellm}.
The industry has been actively exploring deploying LLM services within TEEs as well~\cite{google-private-ai,anthropic-confidential-inference,microsoft-confidential-ai}, with Anthropic recently elaborating on the design principles of confidential inference systems~\cite{anthropic-confidential-inference-systems}.
While our work targets LLM serving within virtualization-based TEEs on the CPU side, it raises a broader question of whether these TEE-protected DNN serving systems are secure against side channel attacks, a direction we leave for future work.

\section{Conclusion}

\noindent
LLMs have traditionally been considered resilient to side-channel attacks due to their largely input-independent execution.
Recent optimizations for LLM serving systems, however, challenge this assumption.
Sparsity-aware optimizations in particular, such as sparse matrix-vector multiplications, introduce input-\emph{dependent} neuron weight accesses.
Our work shows that such dependencies leak neuron activation patterns over memory access side channels,
and that these patterns can be inverted to reconstruct prompt and response tokens with high accuracy,
thereby enabling an end-to-end token extraction attack.
Our attack succeeds even when the optimized LLM serving system is deployed within a TEE (e.g., an Intel~TDX CVM), underscoring the need to carefully assess such optimizations in privacy-sensitive settings.

\section*{Acknowledgment}
\noindent
This material is based upon work supported by Samsung Electronics under grant nr. IO240514\=/09973\=/01, National Research Foundation (NRF) of Korea under grant nr. RS\=/2024\=/00334395, and by Institute of Information \& Communications Technology Planning \& Evaluation (IITP) of the Korea government's Ministry of Science and ICT (MSIT) under grant nr. RS\=/2024\=/00439762.

\IEEEtriggeratref{99}


\bibliographystyle{IEEEtran}
\bibliography{IEEEabrv,biblio}

\begin{thebibliography}{100}
\providecommand{\url}[1]{#1}
\csname url@samestyle\endcsname
\providecommand{\newblock}{\relax}
\providecommand{\bibinfo}[2]{#2}
\providecommand{\BIBentrySTDinterwordspacing}{\spaceskip=0pt\relax}
\providecommand{\BIBentryALTinterwordstretchfactor}{4}
\providecommand{\BIBentryALTinterwordspacing}{\spaceskip=\fontdimen2\font plus
\BIBentryALTinterwordstretchfactor\fontdimen3\font minus \fontdimen4\font\relax}
\providecommand{\BIBforeignlanguage}[2]{{%
\expandafter\ifx\csname l@#1\endcsname\relax
\typeout{** WARNING: IEEEtran.bst: No hyphenation pattern has been}%
\typeout{** loaded for the language `#1'. Using the pattern for}%
\typeout{** the default language instead.}%
\else
\language=\csname l@#1\endcsname
\fi
#2}}
\providecommand{\BIBdecl}{\relax}
\BIBdecl

\bibitem{elhage2022featuresuperposition}
N.~Elhage, T.~Hume, C.~Olsson, N.~Schiefer, T.~Henighan, S.~Kravec, Z.~Hatfield-Dodds, R.~Lasenby, D.~Drain, C.~Chen, R.~Grosse, S.~McCandlish, J.~Kaplan, D.~Amodei, M.~Wattenberg, and C.~Olah, ``Toy models of superposition,'' \emph{arXiv preprint arXiv:2209.10652}, 2022.

\bibitem{bayat2025sparseactivationsteering}
R.~Bayat, A.~Rahimi-Kalahroudi, M.~Pezeshki, S.~Chandar, and P.~Vincent, ``Steering large language model activations in sparse spaces,'' \emph{arXiv preprint arXiv:2503.00177}, 2025.

\bibitem{vaswani2017transformer}
A.~Vaswani, N.~Shazeer, N.~Parmar, J.~Uszkoreit, L.~Jones, A.~N. Gomez, {\L}.~Kaiser, and I.~Polosukhin, ``Attention is all you need,'' in \emph{Advances in Neural Information Processing Systems (NeurIPS)}, 2017.

\bibitem{liu2023dejavu}
Z.~Liu, J.~Wang, T.~Dao, T.~Zhou, B.~Yuan, Z.~Song, A.~Shrivastava, C.~Zhang, Y.~Tian, C.~Re \emph{et~al.}, ``{Deja Vu}: Contextual sparsity for efficient {LLMs} at inference time,'' in \emph{Proceedings of the International Conference on Machine Learning (ICML)}, 2023.

\bibitem{zheng2023pit}
N.~Zheng, H.~Jiang, Q.~Zhang, Z.~Han, L.~Ma, Y.~Yang, F.~Yang, C.~Zhang, L.~Qiu, M.~Yang \emph{et~al.}, ``{PIT}: Optimization of dynamic sparse deep learning models via permutation invariant transformation,'' in \emph{Proceedings of the ACM Symposium on Operating Systems Principles (SOSP)}, 2023.

\bibitem{alizadeh2024llminaflash}
K.~Alizadeh, S.~I. Mirzadeh, D.~Belenko, S.~Khatamifard, M.~Cho, C.~C. Del~Mundo, M.~Rastegari, and M.~Farajtabar, ``{LLM} in a flash: Efficient large language model inference with limited memory,'' in \emph{Proceedings of the Annual Meeting of the Association for Computational Linguistics (ACL)}, 2024.

\bibitem{song2024powerinfer}
Y.~Song, Z.~Mi, H.~Xie, and H.~Chen, ``{PowerInfer}: Fast large language model serving with a consumer-grade {GPU},'' in \emph{Proceedings of the ACM Symposium on Operating Systems Principles (SOSP)}, 2024.

\bibitem{xue2024powerinfer2}
Z.~Xue, Y.~Song, Z.~Mi, L.~Chen, Y.~Xia, and H.~Chen, ``{PowerInfer-2}: Fast large language model inference on a smartphone,'' \emph{arXiv preprint arXiv:2406.06282}, 2024.

\bibitem{van2017withoutpagefaults}
J.~Van~Bulck, N.~Weichbrodt, R.~Kapitza, F.~Piessens, and R.~Strackx, ``Telling your secrets without page faults: Stealthy page table-based attacks on enclaved execution,'' in \emph{Proceedings of the USENIX Security Symposium (Security)}, 2017.

\bibitem{xu2015controlledchannel}
Y.~Xu, W.~Cui, and M.~Peinado, ``Controlled-channel attacks: Deterministic side channels for untrusted operating systems,'' in \emph{Proceedings of the IEEE Symposium on Security and Privacy (IEEE S\&P)}, 2015.

\bibitem{zhuang2004hide}
X.~Zhuang, T.~Zhang, and S.~Pande, ``{HIDE}: An infrastructure for efficiently protecting information leakage on the address bus,'' in \emph{Proceedings of the International Conference on Architectural Support for Programming Languages and Operating Systems (ASPLOS)}, 2004.

\bibitem{lee2020offchip}
D.~Lee, D.~Jung, I.~T. Fang, C.-C. Tsai, and R.~A. Popa, ``An off-chip attack on hardware enclaves via the memory bus,'' in \emph{Proceedings of the USENIX Security Symposium (Security)}, 2020.

\bibitem{chuang2026teefail}
J.~Chuang, A.~Seto, N.~Berrios, S.~van Schaik, C.~Garman, and D.~Genkin, ``Transparent domain extensions: Breaking {Intel} {TEE} implementations via {DDR5} memory bus interposition,'' in \emph{Proceedings of the IEEE Symposium on Security and Privacy (IEEE S\&P)}, 2026, to appear.

\bibitem{seo2024jointentropy}
W.~Seo and J.~Lee, ``Unsupervised feature selection towards pattern discrimination power,'' in \emph{Proceedings of the Conference on Uncertainty in Artificial Intelligence (UAI)}, 2024.

\bibitem{mahendran2015cnninversion}
A.~Mahendran and A.~Vedaldi, ``Understanding deep image representations by inverting them,'' in \emph{Proceedings of the IEEE Conference on Computer Vision and Pattern Recognition (CVPR)}, 2015.

\bibitem{qu2025pia}
W.~Qu, Y.~Zhou, Y.~Wu, T.~Xiao, B.~Yuan, Y.~Li, and J.~Zhang, ``Prompt inversion attack against collaborative inference of large language models,'' in \emph{Proceedings of the IEEE Symposium on Security and Privacy (IEEE S\&P)}, 2025.

\bibitem{dong2025depth}
T.~Dong, Y.~Meng, S.~Li, G.~Chen, Z.~Liu, and H.~Zhu, ``Depth gives a false sense of privacy: {LLM} internal states inversion,'' in \emph{Proceedings of the USENIX Security Symposium (Security)}, 2025.

\bibitem{morris2024lminversion}
J.~X. Morris, W.~Zhao, J.~T. Chiu, V.~Shmatikov, and A.~M. Rush, ``Language model inversion,'' in \emph{Proceedings of the International Conference on Learning Representations (ICLR)}, 2024.

\bibitem{kvm}
\BIBentryALTinterwordspacing
{Open Virtualization Alliance}, ``{Linux} kernel virtual machine.'' [Online]. Available: \url{https://www.linux-kvm.org}
\BIBentrySTDinterwordspacing

\bibitem{bellard2005qemu}
F.~Bellard, ``{QEMU}, a fast and portable dynamic translator,'' in \emph{Proceedings of the USENIX Annual Technical Conference, FREENIX Track}, 2005.

\bibitem{hu2021lora}
E.~J. Hu, Y.~Shen, P.~Wallis, Z.~Allen-Zhu, Y.~Li, S.~Wang, L.~Wang, and W.~Chen, ``{LoRA}: Low-rank adaptation of large language models,'' \emph{arXiv preprint arXiv:2106.09685}, 2021.

\bibitem{li2023transformersparsity}
Z.~Li, C.~You, S.~Bhojanapalli, D.~Li, A.~S. Rawat, S.~J. Reddi, K.~Ye, F.~Chern, F.~Yu, R.~Guo, and S.~Kumar, ``The lazy neuron phenomenon: On emergence of activation sparsity in {Transformers},'' in \emph{Proceedings of the International Conference on Learning Representations (ICLR)}, 2023.

\bibitem{zhang2024relusquared}
Z.~Zhang, Y.~Song, G.~Yu, X.~Han, Y.~Lin, C.~Xiao, C.~Song, Z.~Liu, Z.~Mi, and M.~Sun, ``{ReLU}$^2$ wins: Discovering efficient activation functions for sparse {LLMs},'' \emph{arXiv preprint arXiv:2402.03804}, 2024.

\bibitem{mirzadeh2024relufication}
S.~I. Mirzadeh, K.~Alizadeh-Vahid, S.~Mehta, C.~C. del Mundo, O.~Tuzel, G.~Samei, M.~Rastegari, and M.~Farajtabar, ``{ReLU} strikes back: Exploiting activation sparsity in large language models,'' in \emph{Proceedings of the International Conference on Learning Representations (ICLR)}, 2024.

\bibitem{federici2025dip}
M.~Federici, D.~Belli, M.~V. Baalen, A.~Jalalirad, A.~Skliar, B.~Major, M.~Nagel, and P.~Whatmough, ``Efficient {LLM} inference using dynamic input pruning and cache-aware masking,'' in \emph{Proceedings of the Conference on Machine Learning and Systems (MLSys)}, 2025.

\bibitem{liu2025trainingfreesparsity}
J.~Liu, P.~Ponnusamy, T.~Cai, H.~Guo, Y.~Kim, and B.~Athiwaratkun, ``Training-free activation sparsity in large language models,'' in \emph{Proceedings of the International Conference on Learning Representations (ICLR)}, 2025.

\bibitem{intel-sgx}
\BIBentryALTinterwordspacing
{Intel}, ``Intel\textsuperscript{\textregistered} {Software Guard Extensions} programming reference,'' 2014. [Online]. Available: \url{https://www.intel.com/content/dam/develop/external/us/en/documents/329298-002-629101.pdf}
\BIBentrySTDinterwordspacing

\bibitem{wilke2020sevurity}
L.~Wilke, J.~Wichelmann, M.~Morbitzer, and T.~Eisenbarth, ``{SEVurity}: No security without integrity: Breaking integrity-free memory encryption with minimal assumptions,'' in \emph{Proceedings of the IEEE Symposium on Security and Privacy (IEEE S\&P)}, 2020.

\bibitem{li2019exploitingamdsev}
M.~Li, Y.~Zhang, Z.~Lin, and Y.~Solihin, ``Exploiting unprotected {I/O} operations in {AMD}'s secure encrypted virtualization,'' in \emph{Proceedings of the USENIX Security Symposium (Security)}, 2019.

\bibitem{wilke2024tdxdown}
L.~Wilke, F.~Sieck, and T.~Eisenbarth, ``{TDXdown}: Single-stepping and instruction counting attacks against {Intel TDX},'' in \emph{Proceedings of the ACM Conference on Computer and Communications Security (CCS)}, 2024.

\bibitem{hornetz2026tdxray}
T.~Hornetz, H.~Yavarzadeh, A.~Cheu, A.~Gascon, L.~Gerlach, D.~Moghimi, P.~Schoppmann, M.~Schwarz, and R.~Zhang, ``{TDXRay}: Microarchitectural side-channel analysis of {Intel} {TDX} for real-world workloads,'' in \emph{Proceedings of the IEEE Symposium on Security and Privacy (IEEE S\&P)}, 2026, to appear.

\bibitem{amd-sev}
\BIBentryALTinterwordspacing
{AMD}, ``{AMD SEV-SNP},'' 2020. [Online]. Available: \url{https://www.amd.com/content/dam/amd/en/documents/epyc-business-docs/white-papers/SEV-SNP-strengthening-vm-isolation-with-integrity-protection-and-more.pdf}
\BIBentrySTDinterwordspacing

\bibitem{intel-tdx}
\BIBentryALTinterwordspacing
{Intel}, ``Intel\textsuperscript{\textregistered} {Trust Domain Extensions},'' 2022. [Online]. Available: \url{https://cdrdv2-public.intel.com/690419/TDX-Whitepaper-February2022.pdf}
\BIBentrySTDinterwordspacing

\bibitem{intel-tdx-module-spec}
\BIBentryALTinterwordspacing
------, ``Intel\textsuperscript{\textregistered} {Trust Domain Extensions} ({Intel}\textsuperscript{\textregistered} {TDX}) module base architecture specification,'' 2025. [Online]. Available: \url{https://cdrdv2-public.intel.com/853286/intel-tdx-module-base-spec-348549006.pdf}
\BIBentrySTDinterwordspacing

\bibitem{qemuuserguide}
\BIBentryALTinterwordspacing
``{QEMU} system emulation user's guide.'' [Online]. Available: \url{https://www.qemu.org/docs/master/system/index.html}
\BIBentrySTDinterwordspacing

\bibitem{mo2020darknetz}
F.~Mo, A.~S. Shamsabadi, K.~Katevas, S.~Demetriou, I.~Leontiadis, A.~Cavallaro, and H.~Haddadi, ``{DarkneTZ}: Towards model privacy at the edge using trusted execution environments,'' in \emph{Proceedings of the Annual International Conference on Mobile Systems, Applications, and Services (MobiSys)}, 2020.

\bibitem{siby2024guarantee}
S.~Siby, S.~Abdollahi, M.~Maheri, M.~Kogias, and H.~Haddadi, ``{GuaranTEE}: Towards attestable and private {ML} with {CCA},'' in \emph{Proceedings of the Workshop on Machine Learning and Systems (EuroMLSys)}, 2024.

\bibitem{moon2025asgard}
M.~Moon, M.~Kim, J.~Jung, and D.~Song, ``{ASGARD}: Protecting on-device deep neural networks with virtualization-based trusted execution environments,'' in \emph{Proceedings of the Network and Distributed System Security Symposium (NDSS)}, 2025.

\bibitem{tan2025pipellm}
Y.~Tan, C.~Tan, Z.~Mi, and H.~Chen, ``{PipeLLM}: Fast and confidential large language model services with speculative pipelined encryption,'' in \emph{Proceedings of the International Conference on Architectural Support for Programming Languages and Operating Systems (ASPLOS)}, 2025.

\bibitem{wang2026tzllm}
X.~Wang, J.~Shi, Z.~Zhao, Y.~Yu, Z.~Hua, and J.~Gu, ``{TZ-LLM}: Protecting on-device large language models with {Arm TrustZone},'' in \emph{Proceedings of the ACM European Conference on Computer Systems (EuroSys)}, 2026.

\bibitem{google-private-ai}
\BIBentryALTinterwordspacing
{Google}, ``Enabling more private generative {AI},'' 2024. [Online]. Available: \url{https://developers.googleblog.com/en/enabling-more-private-gen-ai}
\BIBentrySTDinterwordspacing

\bibitem{anthropic-confidential-inference}
\BIBentryALTinterwordspacing
{Anthropic}, ``Confidential inference via trusted virtual machines,'' 2025. [Online]. Available: \url{https://www.anthropic.com/research/confidential-inference-trusted-vms}
\BIBentrySTDinterwordspacing

\bibitem{microsoft-confidential-ai}
\BIBentryALTinterwordspacing
{Microsoft}, ``Confidential {AI},'' 2023. [Online]. Available: \url{https://learn.microsoft.com/en-us/azure/confidential-computing/confidential-ai}
\BIBentrySTDinterwordspacing

\bibitem{chrapek2025confidentialllminference}
M.~Chrapek, M.~Copik, E.~Mettaz, and T.~Hoefler, ``Confidential {LLM} inference: Performance and cost across {CPU} and {GPU} {TEEs},'' in \emph{Proceedings of the IEEE International Symposium on Workload Characterization (IISWC)}, 2025.

\bibitem{shoeybi2019megatron}
M.~Shoeybi, M.~Patwary, R.~Puri, P.~LeGresley, J.~Casper, and B.~Catanzaro, ``{Megatron-LM}: Training multi-billion parameter language models using model parallelism,'' \emph{arXiv preprint arXiv:1909.08053}, 2019.

\bibitem{wu2025promptpeek}
Y.~Z. Guanlong~Wu, Zheng~Zhang, ``I know what you asked: Prompt leakage via {KV}-cache sharing in multi-tenant {LLM} serving,'' in \emph{Proceedings of the Network and Distributed System Security Symposium (NDSS)}, 2025.

\bibitem{gao2025iknowwhatyousaid}
Z.~Gao, J.~Hu, F.~Guo, Y.~Zhang, Y.~Han, S.~Liu, H.~Li, and Z.~Lv, ``I know what you said: Unveiling hardware cache side-channels in local large language model inference,'' in \emph{Proceedings of the USENIX Security Symposium (Security)}, 2025.

\bibitem{kang2021deephashembedding}
W.-C. Kang, D.~Z. Cheng, T.~Yao, X.~Yi, T.~Chen, L.~Hong, and E.~H. Chi, ``Learning to embed categorical features without embedding tables for recommendation,'' in \emph{Proceedings of the ACM SIGKDD Conference on Knowledge Discovery \& Data Mining}, 2021.

\bibitem{umar2025deephashembedding}
M.~Umar, A.~P. Marathe, M.~D. Gupta, S.~J. Ghosh, G.~E. Suh, and W.~Xiong, ``Efficient memory side-channel protection for embedding generation in machine learning,'' in \emph{Proceedings of the IEEE International Symposium on High Performance Computer Architecture (HPCA)}, 2025.

\bibitem{russel2008virtio}
R.~Russell, ``{virtio}: Towards a de-facto standard for virtual {I/O} devices,'' \emph{SIGOPS Operating Systems Review}, vol.~42, no.~5, pp. 95--103, Jul. 2008.

\bibitem{virtiov1.2}
\BIBentryALTinterwordspacing
``Virtual {I/O} device ({VIRTIO}) version 1.2,'' 2022. [Online]. Available: \url{https://docs.oasis-open.org/virtio/virtio/v1.2/virtio-v1.2.pdf}
\BIBentrySTDinterwordspacing

\bibitem{google2024dmverity}
\BIBentryALTinterwordspacing
{Google}, ``Implementing dm-verity,'' 2024. [Online]. Available: \url{https://source.android.com/docs/security/features/verifiedboot/dm-verity}
\BIBentrySTDinterwordspacing

\bibitem{intel-tdx-google-review}
\BIBentryALTinterwordspacing
E.~Aktas, C.~Cohen, J.~Eads, J.~Forshaw, and F.~Wilhelm, ``{Intel} trust domain extensions ({TDX}) security review,'' 2023. [Online]. Available: \url{https://services.google.com/fh/files/misc/intel_tdx_-_full_report_041423.pdf}
\BIBentrySTDinterwordspacing

\bibitem{luo2026kvinversion}
Z.~Luo, S.~Shao, S.~Zhang, L.~Zhou, Y.~Hu, C.~Zhao, Z.~Liu, and Z.~Qin, ``Shadow in the cache: Unveiling and mitigating privacy risks of {KV}-cache in {LLM} inference,'' in \emph{Proceedings of the Network and Distributed System Security Symposium (NDSS)}, 2026.

\bibitem{pasquini2025llmmap}
D.~Pasquini, E.~M. Kornaropoulos, and G.~Ateniese, ``{LLMmap}: Fingerprinting for large language models,'' in \emph{Proceedings of the USENIX Security Symposium (Security)}, 2025.

\bibitem{hua2018reversecnn}
W.~Hua, Z.~Zhang, and G.~E. Suh, ``Reverse engineering convolutional neural networks through side-channel information leaks,'' in \emph{Proceedings of the Annual Design Automation Conference (DAC)}, 2018.

\bibitem{yan2020cachetelepathy}
M.~Yan, C.~W. Fletcher, and J.~Torrellas, ``Cache telepathy: Leveraging shared resource attacks to learn {DNN} architectures,'' in \emph{Proceedings of the USENIX Security Symposium (Security)}, 2020.

\bibitem{gao2024deeptheft}
Y.~Gao, H.~Qiu, Z.~Zhang, B.~Wang, H.~Ma, A.~Abuadbba, M.~Xue, A.~Fu, and S.~Nepal, ``{DeepTheft}: Stealing {DNN} model architectures through power side channel,'' in \emph{Proceedings of the IEEE Symposium on Security and Privacy (IEEE S\&P)}, 2024.

\bibitem{llama2}
H.~Touvron, L.~Martin, K.~Stone, P.~Albert, A.~Almahairi, Y.~Babaei, N.~Bashlykov, S.~Batra, P.~Bhargava, S.~Bhosale \emph{et~al.}, ``Llama 2: Open foundation and fine-tuned chat models,'' \emph{arXiv preprint arXiv:2307.09288}, 2023.

\bibitem{rauscher2025tdxploit}
F.~Rauscher, L.~Wilke, H.~Weissteiner, T.~Eisenbarth, and D.~Gruss, ``{TDXploit}: Novel techniques for single-stepping and cache attacks on {Intel} {TDX},'' in \emph{Proceedings of the USENIX Security Symposium (Security)}, 2025.

\bibitem{downey2008littlebookofsemaphores}
A.~Downey, \emph{The little book of semaphores}.\hskip 1em plus 0.5em minus 0.4em\relax Green Tea Press, 2008, vol.~2, no.~2.

\bibitem{he2025lawnexttoken}
H.~He and W.~J. Su, ``A law of next-token prediction in large language models,'' \emph{Phys. Rev. E}, 2025.

\bibitem{zhang2022opt}
S.~Zhang, S.~Roller, N.~Goyal, M.~Artetxe, M.~Chen, S.~Chen, C.~Dewan, M.~Diab, X.~Li, X.~V. Lin, T.~Mihaylov, M.~Ott, S.~Shleifer, K.~Shuster, D.~Simig, P.~S. Koura, A.~Sridhar, T.~Wang, and L.~Zettlemoyer, ``{OPT}: Open pre-trained {Transformer} language models,'' 2022.

\bibitem{sparsellm2023sparsellm}
\BIBentryALTinterwordspacing
{SpaseLLM Team}, ``Sparse large language models with {ReLU} activation,'' 2023. [Online]. Available: \url{https://huggingface.co/SparseLLM/ReluLLaMA-7B}
\BIBentrySTDinterwordspacing

\bibitem{nvidia2023nemotron}
\BIBentryALTinterwordspacing
{NVIDIA}, ``{Nemotron-3-8B-Base-4k},'' 2023. [Online]. Available: \url{https://huggingface.co/nvidia/nemotron-3-8b-base-4k}
\BIBentrySTDinterwordspacing

\bibitem{touvron2023llama2openfoundation}
H.~Touvron, L.~Martin, K.~Stone, P.~Albert, A.~Almahairi, Y.~Babaei, N.~Bashlykov, S.~Batra, P.~Bhargava, S.~Bhosale, D.~Bikel, L.~Blecher, C.~C. Ferrer, M.~Chen, G.~Cucurull, D.~Esiobu, J.~Fernandes, J.~Fu, W.~Fu, B.~Fuller, C.~Gao, V.~Goswami, N.~Goyal, A.~Hartshorn, S.~Hosseini, R.~Hou, H.~Inan, M.~Kardas, V.~Kerkez, M.~Khabsa, I.~Kloumann, A.~Korenev, P.~S. Koura, M.-A. Lachaux, T.~Lavril, J.~Lee, D.~Liskovich, Y.~Lu, Y.~Mao, X.~Martinet, T.~Mihaylov, P.~Mishra, I.~Molybog, Y.~Nie, A.~Poulton, J.~Reizenstein, R.~Rungta, K.~Saladi, A.~Schelten, R.~Silva, E.~M. Smith, R.~Subramanian, X.~E. Tan, B.~Tang, R.~Taylor, A.~Williams, J.~X. Kuan, P.~Xu, Z.~Yan, I.~Zarov, Y.~Zhang, A.~Fan, M.~Kambadur, S.~Narang, A.~Rodriguez, R.~Stojnic, S.~Edunov, and T.~Scialom, ``Llama 2: Open foundation and fine-tuned chat models,'' \emph{arXiv preprint arXiv:2307.09288}, 2023.

\bibitem{google2024gemma}
{Gemma Team}, ``{Gemma}: Open models based on {Gemini} research and technology,'' \emph{arXiv preprint arXiv:2403.08295}, 2024.

\bibitem{danisman2019skytraxairlinereviews}
\BIBentryALTinterwordspacing
E.~Danisman, ``Skytrax airline reviews,'' 2019. [Online]. Available: \url{https://www.kaggle.com/datasets/efehandanisman/skytrax-airline-reviews}
\BIBentrySTDinterwordspacing

\bibitem{han2023medalpaca}
T.~Han, L.~C. Adams, J.-M. Papaioannou, P.~Grundmann, T.~Oberhauser, A.~L{\"o}ser, D.~Truhn, and K.~K. Bressem, ``{MedAlpaca}--an open-source collection of medical conversational {AI} models and training data,'' \emph{arXiv preprint arXiv:2304.08247}, 2023.

\bibitem{chalkidis2019neurallegaljudgmentprediction}
\BIBentryALTinterwordspacing
I.~Chalkidis, I.~Androutsopoulos, and N.~Aletras, ``Neural legal judgment prediction in {English},'' 2019. [Online]. Available: \url{https://arxiv.org/abs/1906.02059}
\BIBentrySTDinterwordspacing

\bibitem{morris2023privateprompts}
\BIBentryALTinterwordspacing
J.~Morris, ``Private prompts,'' 2023. [Online]. Available: \url{https://huggingface.co/datasets/jxm/private_prompts}
\BIBentrySTDinterwordspacing

\bibitem{chua2024systempromptleakage}
\BIBentryALTinterwordspacing
G.~Chua, ``System prompt leakage,'' 2024. [Online]. Available: \url{https://huggingface.co/datasets/gabrielchua/system-prompt-leakage}
\BIBentrySTDinterwordspacing

\bibitem{zhang2024output2prompt}
C.~Zhang, J.~X. Morris, and V.~Shmatikov, ``Extracting prompts by inverting {LLM} outputs,'' in \emph{Proceedings of the Conference on Empirical Methods in Natural Language Processing (EMNLP)}, 2024.

\bibitem{song2025earlybird}
L.~Song, Z.~Pang, W.~Wang, Z.~Wang, X.~Wang, H.~Chen, W.~Song, Y.~Jin, D.~Meng, and R.~Hou, ``The early bird catches the leak: Unveiling timing side channels in {LLM} serving systems,'' \emph{arXiv preprint arXiv:2409.20002}, 2025.

\bibitem{das2025systempromptextraction}
B.~C. Das, M.~H. Amini, and Y.~Wu, ``System prompt extraction attacks and defenses in large language models,'' \emph{arXiv preprint arXiv:2505.23817}, 2025.

\bibitem{salemi2024lamp}
A.~Salemi, S.~Mysore, M.~Bendersky, and H.~Zamani, ``{LaMP}: When large language models meet personalization,'' in \emph{Proceedings of the Annual Meeting of the Association for Computational Linguistics (ACL)}, 2024.

\bibitem{tan2024democratizing}
Z.~Tan, Q.~Zeng, Y.~Tian, Z.~Liu, B.~Yin, and M.~Jiang, ``Democratizing large language models via personalized parameter-efficient fine-tuning,'' in \emph{Proceedings of the Conference on Empirical Methods in Natural Language Processing (EMNLP)}, 2024.

\bibitem{chen2024persona}
J.~Chen, X.~Wang, R.~Xu, S.~Yuan, Y.~Zhang, W.~Shi, J.~Xie, S.~Li, R.~Yang, T.~Zhu \emph{et~al.}, ``From persona to personalization: A survey on role-playing language agents,'' \emph{arXiv preprint arXiv:2404.18231}, 2024.

\bibitem{xu2025personalized}
Y.~Xu, J.~Zhang, A.~Salemi, X.~Hu, W.~Wang, F.~Feng, H.~Zamani, X.~He, and T.-S. Chua, ``Personalized generation in large model era: A survey,'' in \emph{Proceedings of the Annual Meeting of the Association for Computational Linguistics (ACL)}, 2025.

\bibitem{llama.cpp}
\BIBentryALTinterwordspacing
G.~Gerganov, ``{llama.cpp}: {LLM} inference in {C/C++},'' 2025. [Online]. Available: \url{https://github.com/ggml-org/llama.cpp}
\BIBentrySTDinterwordspacing

\bibitem{tracepoint}
\BIBentryALTinterwordspacing
``Using the {Linux} kernel {Tracepoints}.'' [Online]. Available: \url{https://www.kernel.org/doc/Documentation/trace/tracepoints.txt}
\BIBentrySTDinterwordspacing

\bibitem{ebpf}
\BIBentryALTinterwordspacing
B.~Gregg, ``{Linux} extended {BPF} ({eBPF}) tracing tools,'' 2018. [Online]. Available: \url{http://www.brendangregg.com/ebpf.html}
\BIBentrySTDinterwordspacing

\bibitem{papineni2002bleu}
K.~Papineni, S.~Roukos, T.~Ward, and W.-J. Zhu, ``{BLEU}: a method for automatic evaluation of machine translation,'' in \emph{Proceedings of the Annual Meeting of the Association for Computational Linguistics (ACL)}, 2002.

\bibitem{wikidump}
\BIBentryALTinterwordspacing
{Wikimedia Foundation}. Wikimedia downloads. [Online]. Available: \url{https://dumps.wikimedia.org}
\BIBentrySTDinterwordspacing

\bibitem{dmcrypt67}
\BIBentryALTinterwordspacing
{The kernel development community}, ``dm-crypt,'' 2024. [Online]. Available: \url{https://www.kernel.org/doc/html/v6.7/admin-guide/device-mapper/dm-crypt.html}
\BIBentrySTDinterwordspacing

\bibitem{goldreich1996oram}
O.~Goldreich and R.~Ostrovsky, ``Software protection and simulation on oblivious {RAMs},'' \emph{Journal of the ACM (JACM)}, vol.~43, no.~3, pp. 431--473, 1996.

\bibitem{rane2015raccoon}
A.~Rane, C.~Lin, and M.~Tiwari, ``Raccoon: Closing digital side-channels through obfuscated execution,'' in \emph{Proceedings of the USENIX Security Symposium (Security)}, 2015.

\bibitem{nayak2017hop}
K.~Nayak, C.~W. Fletcher, L.~Ren, N.~Chandran, S.~V. Lokam, E.~Shi, and V.~Goyal, ``{HOP}: Hardware makes obfuscation practical,'' in \emph{Proceedings of the Network and Distributed System Security Symposium (NDSS)}, 2017.

\bibitem{sheng2023flexgen}
Y.~Sheng, L.~Zheng, B.~Yuan, Z.~Li, M.~Ryabinin, B.~Chen, P.~Liang, C.~R\'{e}, I.~Stoica, and C.~Zhang, ``{FlexGen}: high-throughput generative inference of large language models with a single {GPU},'' in \emph{Proceedings of the International Conference on Machine Learning (ICML)}, 2023.

\bibitem{lee2024infinigen}
W.~Lee, J.~Lee, J.~Seo, and J.~Sim, ``{InfiniGen}: Efficient generative inference of large language models with dynamic {KV} cache management,'' in \emph{Proceedings of the USENIX Symposium on Operating Systems Design and Implementation (OSDI)}, 2024.

\bibitem{park2024improving}
D.~Park and B.~Egger, ``Improving throughput-oriented {LLM} inference with {CPU} computations,'' in \emph{Proceedings of the International Conference on Parallel Architectures and Compilation Techniques (PACT)}, 2024.

\bibitem{liu2024deepcache}
Z.~Liu, Y.~Yuan, Y.~Chen, S.~Hu, T.~Li, and S.~Wang, ``{DeepCache}: Revisiting cache side-channel attacks in deep neural networks executables,'' in \emph{Proceedings of the ACM Conference on Computer and Communications Security (CCS)}, 2024.

\bibitem{maas2013phantom}
M.~Maas, E.~Love, E.~Stefanov, M.~Tiwari, E.~Shi, K.~Asanovic, J.~Kubiatowicz, and D.~Song, ``Phantom: Practical oblivious computation in a secure processor,'' in \emph{Proceedings of the ACM Conference on Computer and Communications Security (CCS)}, 2013.

\bibitem{yarom2014flushreload}
Y.~Yarom and K.~Falkner, ``{FLUSH+RELOAD}: A high resolution, low noise, {L3} cache side-channel attack,'' in \emph{Proceedings of the USENIX Security Symposium (Security)}, 2014.

\bibitem{gruss2016flushflush}
D.~Gruss, C.~Maurice, K.~Wagner, and S.~Mangard, ``{Flush+Flush}: a fast and stealthy cache attack,'' in \emph{Proceedings of the International Conference on Detection of Intrusions and Malware, and Vulnerability Assessment (DIMVA)}, 2016.

\bibitem{spielman2025controlledmodelstealing}
J.~Spielman, D.~Oswald, M.~Ryan, and J.~Van~Bulck, ``Activation functions considered harmful: Recovering neural network weights through controlled channels,'' in \emph{Proceedings of the International Symposium on Research in Attacks, Intrusions and Defenses (RAID)}, 2025.

\bibitem{yuan2024hypertheft}
Y.~Yuan, Z.~Liu, S.~Deng, Y.~Chen, S.~Wang, Y.~Zhang, and Z.~Su, ``{HyperTheft}: Thieving model weights from {TEE}-shielded neural networks via ciphertext side channels,'' in \emph{Proceedings of the ACM Conference on Computer and Communications Security (CCS)}, 2024.

\bibitem{gongye2020floatingpoint}
C.~Gongye, Y.~Fei, and T.~Wahl, ``Reverse-engineering deep neural networks using floating-point timing side-channels,'' in \emph{Proceedings of the ACM/EDAC/IEEE Design Automation Conference (DAC)}, 2020.

\bibitem{yan2025relocatevote}
Y.~Yan, W.~Huang, I.~Grishchenko, G.~Saileshwar, A.~Mehta, and D.~Lie, ``{Relocate-Vote}: Using sparsity information to exploit ciphertext side-channels,'' in \emph{Proceedings of the USENIX Security Symposium (Security)}, 2025.

\bibitem{zheng2024inputsnatch}
X.~Zheng, H.~Han, S.~Shi, Q.~Fang, Z.~Du, Q.~Guo, and X.~Hu, ``{InputSnatch}: Stealing input in {LLM} services via timing side-channel attacks,'' \emph{arXiv preprint arXiv:2411.18191}, 2024.

\bibitem{zhang2022moefication}
Z.~Zhang, Y.~Lin, Z.~Liu, P.~Li, M.~Sun, and J.~Zhou, ``{M}o{E}fication: {Transformer} feed-forward layers are mixtures of experts,'' in \emph{Findings of the Association for Computational Linguistics: ACL 2022}, 2022.

\bibitem{szatkowski2024d2dmoefication}
F.~Szatkowski, B.~W{\'o}jcik, M.~Pi{\'o}rczy{\'n}ski, and S.~Scardapane, ``Exploiting activation sparsity with dense to dynamic-k mixture-of-experts conversion,'' in \emph{Advances in Neural Information Processing Systems (NeurIPS)}, 2024.

\bibitem{hayes2024bufferovfmoe}
J.~Hayes, I.~Shumailov, and I.~Yona, ``Buffer overflow in mixture of experts,'' \emph{arXiv preprint arXiv:2402.05526}, 2024.

\bibitem{yona2024promptstealingfrommoe}
I.~Yona, I.~Shumailov, J.~Hayes, and N.~Carlini, ``Stealing user prompts from mixture of experts,'' \emph{arXiv preprint arXiv:2410.22884}, 2024.

\bibitem{ding2025moecho}
R.~Ding, T.~Xu, X.~Shen, A.~A. Ding, and Y.~Fei, ``{MoEcho}: Exploiting side-channel attacks to compromise user privacy in mixture-of-experts {LLMs},'' in \emph{Proceedings of the ACM Conference on Computer and Communications Security (CCS)}, 2025.

\bibitem{leviathan2023specdecode}
Y.~Leviathan, M.~Kalman, and Y.~Matias, ``Fast inference from {Transformers} via speculative decoding,'' in \emph{Proceedings of the International Conference on Machine Learning (ICML)}, 2023.

\bibitem{wei2024specdecodingidechannel}
J.~Wei, A.~Abdulrazzag, T.~Zhang, A.~Muursepp, and G.~Saileshwar, ``Privacy risks of speculative decoding in large language models,'' \emph{arXiv preprint arXiv:2411.01076}, 2024.

\bibitem{zhang2024systempromptextraction}
Y.~Zhang, N.~Carlini, and D.~Ippolito, ``Effective prompt extraction from language models,'' in \emph{Proceedings of the Conference on Language Modeling (COLM)}, 2024.

\bibitem{hui2024pleak}
B.~Hui, H.~Yuan, N.~Gong, P.~Burlina, and Y.~Cao, ``{PLeak}: Prompt leaking attacks against large language model applications,'' in \emph{Proceedings of the ACM Conference on Computer and Communications Security (CCS)}, 2024.

\bibitem{mo2024mlwithcc}
F.~Mo, Z.~Tarkhani, and H.~Haddadi, ``Machine learning with confidential computing: A systematization of knowledge,'' \emph{ACM Computing Survey (CSUR)}, vol.~56, no.~11, Jun. 2024.

\bibitem{sun2022shadownet}
Z.~Sun, R.~Sun, C.~Liu, A.~R. Chowdhury, L.~Lu, and S.~Jha, ``{ShadowNet}: A secure and efficient on-device model inference system for convolutional neural networks,'' in \emph{Proceedings of the IEEE Symposium on Security and Privacy (IEEE S\&P)}, 2023.

\bibitem{zhang2024teeslice}
Z.~Zhang, C.~Gong, Y.~Cai, Y.~Yuan, B.~Liu, D.~Li, Y.~Guo, and X.~Chen, ``No privacy left outside: On the (in-)security of {TEE}-shielded {DNN} partition for on-device {ML},'' in \emph{Proceedings of the IEEE Symposium on Security and Privacy (IEEE S\&P)}, 2024.

\bibitem{shen2022soter}
T.~Shen, J.~Qi, J.~Jiang, X.~Wang, S.~Wen, X.~Chen, S.~Zhao, S.~Wang, L.~Chen, X.~Luo, F.~Zhang, and H.~Cui, ``{SOTER}: Guarding black-box inference for general neural networks at the edge,'' in \emph{Proceedings of the USENIX Annual Technical Conference (ATC)}, 2022.

\bibitem{lee2019occlumency}
T.~Lee, Z.~Lin, S.~Pushp, C.~Li, Y.~Liu, Y.~Lee, F.~Xu, C.~Xu, L.~Zhang, and J.~Song, ``{Occlumency}: Privacy-preserving remote deep-learning inference using {SGX},'' in \emph{Proceedings of the Annual International Conference on Mobile Computing and Networking (MobiCom)}, 2019.

\bibitem{feng2024snpu}
E.~Feng, D.~Feng, D.~Du, Y.~Xia, and H.~Chen, ``{sNPU}: Trusted execution environments on integrated {NPUs},'' in \emph{Proceedings of the ACM/IEEE Annual International Symposium on Computer Architecture (ISCA)}, 2024.

\bibitem{nvidia-h100}
\BIBentryALTinterwordspacing
{NVIDIA}, ``Confidential {Compute on {NVIDIA} {Hopper} {H100}},'' 2023. [Online]. Available: \url{https://images.nvidia.com/aem-dam/en-zz/Solutions/data-center/HCC-Whitepaper-v1.0.pdf}
\BIBentrySTDinterwordspacing

\bibitem{anthropic-confidential-inference-systems}
\BIBentryALTinterwordspacing
{Pattern Labs} and {Anthropic}, ``Confidential inference systems: Design principles and security risks,'' 2025. [Online]. Available: \url{https://assets.anthropic.com/m/c52125297b85a42/original/Confidential_Inference_Paper.pdf}
\BIBentrySTDinterwordspacing

\end{thebibliography}

\end{document}